\documentclass[
 pra, onecolumn,
 superscriptaddress,
 amsmath,
 amssymb,
 aps
]{revtex4-2}

\usepackage[colorlinks = true,
            linkcolor = blue,
            urlcolor  = blue,
            citecolor = blue,
            anchorcolor = blue]{hyperref}

\usepackage[english]{babel}
\usepackage{latexsym}
\usepackage{graphics}
\usepackage{graphicx}
\usepackage{epsfig}
\usepackage{color}
\usepackage{bm}
\usepackage{amssymb}
\usepackage{amsmath}
\usepackage{amsthm}
\usepackage{dcolumn}
\usepackage{float}
\usepackage{epstopdf}
\usepackage{cleveref}
\usepackage[svgnames]{xcolor}
\usepackage{tabularx}
\usepackage{mathtools}
\usepackage{listings}

\newcolumntype{L}{>{\raggedright\arraybackslash}X}

\DeclarePairedDelimiter{\ceil}{\lceil}{\rceil}
\theoremstyle{remark}

\usepackage{braket}
\usepackage{wrapfig}
\usepackage{caption}
\usepackage{dcolumn}

\usepackage[flushleft]{threeparttable}
\usepackage{tablefootnote}

\usepackage{multirow}
\usepackage{makecell}

\makeatletter
\renewcommand{\p@subsection}{}
\renewcommand{\p@subsubsection}{}

\def\@cline#1-#2\@nil{%
\omit
\@multicnt#1%
\advance\@multispan\m@ne
\ifnum\@multicnt=\@ne\@firstofone{&\omit}\fi
\@multicnt#2%
\advance\@multicnt-#1%
\advance\@multispan\@ne
\leaders\hrule\@height\arrayrulewidth\hfill
\cr
\noalign{\nobreak\vskip-\arrayrulewidth}}

\makeatother

\begin{document}
    \title{Arline Benchmarks: Automated Benchmarking Platform for Quantum Compilers}
    \author{Y. Kharkov} 
    \email[]{info@arline.io}
    \author{A. Ivanova}
    \author{E. Mikhantiev}
    \author{A. Kotelnikov}
    
    \affiliation{Arline}

    \date{\today}
    
    \begin{abstract}
        Efficient compilation of quantum algorithms is vital in the era of Noisy Intermediate-Scale Quantum (NISQ) devices. While multiple open-source quantum compilation and circuit optimization frameworks are available, e.g. IBM Qiskit, CQC Tket, Google Cirq, Rigetti Quilc, PyZX, their relative performance  is not always clear to a quantum programmer.
        The growth of complexity and diversity of quantum circuit compilation algorithms creates a demand for a dedicated tool for cross-benchmarking and profiling of inner workflow of the quantum compilation stack.  
        We present an open-source software package, Arline Benchmarks, that is designed to perform automated benchmarking of quantum compilers with the focus on NISQ applications. The name ``Arline'' was given in honour of Arline Greenbaum Feynman, the first wife of Richard Feynman, the pioneer of quantum computing.
        We compared several quantum compilation frameworks  based on a set of important metrics such as gate counts, circuit depth, hardware-dependent circuit cost function, compiler run time etc. with a detailed analysis of metrics for each compilation stage.
        We executed a variety of compiler tests for random  circuits and  structured quantum algorithms (VQE, Trotter decomposition, Grover search, Option Pricing via Amplitude Estimation) for several popular quantum hardware architectures.
        We also compare compilers' metrics with theoretical benchmarks for two-qubit (KAK) and three-qubit circuits.
        In addition, by leveraging a versatile set of compilation subroutines, Arline platform allows to achieve an additional improvement in circuit compression, when compared to the performance of compilers from individual vendors. 
        We  propose a concept of composite compilation pipeline that combines compiler-specific circuit optimization subroutines in a single compilation stack to find an optimized sequence of compilation passes.
    By providing detailed insights into the compilation flow of quantum compilers, Arline Benchmarks offers a valuable toolkit for quantum computing researchers and software developers to gain additional insights into compilers' characteristics.         \end{abstract}

    \maketitle

    \section{Introduction}

    Quantum compilation is a problem of translating a quantum algorithm in to a set of low-level hardware instructions to be executed on a quantum processor.
    Modern quantum compilers perform  circuit optimizations prior to execution on a hardware aiming to minimize the number of gates in a quantum algorithm. 
    By the means of optimizing gate count in a quantum circuit, it is possible to significantly reduce  hardware errors thus raising the barriers for building large scale quantum computers.
    Progress towards practical quantum computing will require significant effort in building hardware-software co-design workflows in which efficient quantum compilation pipelines will play crucial role~\cite{chong2017programming}.
    
    Quantum compilation software frameworks are quickly evolving and becoming increasingly complex supported by significant research effort for designing new circuit optimization algorithms.
    Some of the most popular quantum compilation frameworks with inbuild circuit optimization functionality include open-source libraries IBM Qiskit~\cite{qiskit}, Google Cirq~\cite{cirq}, Rigetti Quilc~\cite{quilc2020}, PyZX library~\cite{kissinger2020Pyzx} based on ZX-calculus and Tket compiler from Cambridge Quantum Computing~\cite{pytket}.
    Each of the aforementioned frameworks contains a  set of specific compilation subroutines, which have their own advantages and constraints depending on the properties of the input circuit and quantum hardware. Thus, detailed benchmarking  of quantum compilation workflows for near-term quantum algorithms is  crucial for understanding interactions between various compilation modules as well as for improving/debugging  quantum compilers.
    
    A comprehensive evaluation of modern quantum compilers  requires expert-level knowledge of compilation workflow, quantum algorithms and quantum hardware. The complexity of quantum compiler benchmarking is worsened by diversity of competing quantum hardware platforms (superconducting qubits, trapped ions, Rydberg atoms, photonic quantum processors), since each of the platforms have a specific hardware-native gate set, qubit connectivity and noise characteristics that makes it difficult to define a single cost metric. Hence, benchmarking results should be assessed based on a variety of carefully designed tests and relevant metrics. 
    It is interesting to note, that benchmarking problems of a similar level of complexity arise in the context of machine learning, where cross-benchmarking of different models and hardware is needed.  This problem inspired machine leearning community to create MLPerf project~\cite{mattson2020mlperf}, with the goal of creating fair and useful benchmarks representing the state of the art in artificial intelligence.
    Arline Benchmarks project has a similar ambitious goal for establishing the state of the art in performance of modern quantum compilers. We should also mention QASMBench, which is a recent effort in establishing benchmarking suits (circuit datasets) for evaluation of NISQ devices~\cite{li2020qasmbench}, that relies on  OpenQASM assembly-level representation of quantum circuits~\cite{cross2017open}. 
    
    Previous empirical studies of quantum compilers 
    \cite{tket2020, quilc2020, murali2019full, tan2020optimal} compared performance of several compilation frameworks on domain-specific classes of quantum circuits. 
    Meanwhile, the understanding of compilers' circuit optimization performance remains quite limited since it is difficult to predict how effective will be a certain sequence of optimization passes for a specific quantum circuit.  Different compilation subroutines applied sequentially can have a complex  effect on each other and result in hardly predictable outcome for the chosen metric of interest, e.g. two-qubit gate count of circuit depth. 
    The cross-compiler functionality of Arline platform allows combining subroutines from  different vendors with the aim of designing more optimal compilation stacks. 
    We consider the interplay of circuit optimization subroutines and showed that by searching through a large space of possible combinations of passes, one can discover more efficient sequences of compilation passes.   
    
    Most importantly, end-to-end software solutions which would allow to streamline benchmarking processes are still lacking. 
    Arline Benchmarks platform aims to provide a fair comparison between compilation frameworks for various  quantum hardware and quantum algorithms with a focus on NISQ applications. 
    Our platform allows to visualize  a transformation of circuit metrics throughout the compilation pipeline and pinpoint strengths and weaknesses of individual compilers' subroutines.
    Our benchmarking experiments clearly show that the performance of  circuit optimization pipelines is strongly dependent on the choice of the class of target circuits as well as  quantum hardware configuration.
    As the crucial metrics for accessing the performance of compilers we use the two-qubit gate counts/depth as well as we utilize an average circuit log-fidelity as a heuristic cost function.

    The structure of the paper goes as follows. In Sec. \ref{sec:platform} we describe the main components of the Arline Benchmarks platform and Arline's i/o interface, provide relevant circuit metrics for the benchmarking, and give a short overview of the common circuit compilation subroutines. 
    Next, in Sec. \ref{sec:kak+3q} we 
    consider optimization of two-qubit and three-qubit random circuits with open-source compilers and compare results with theoretically optimal bounds.
    Benchmarking of routing/mapping subroutines for hardware with restricted qubit connectivity is performed in Sec. \ref{sec:routing}. Sec. \ref{sec:compression}
    contains benchmarking results for random and structured target circuits on hardware with full connectivity as well as on several popular hardware architectures.
    In Sec. \ref{sec:ranking} we test circuit optimization subroutines in isolation and provide a ranking 
    of subroutines according to their circuit optimization performance. We present one of the most intriguing results of the paper in Sec. \ref{sec:composite}, where we show that by combining compilation subroutines from different software vendors in a specific sequential order it is possible to achieve superior circuit optimization performance.
    Finally, we summarize our results in Sec. \ref{sec:conclusions}  and discuss future potential extensions of Arline platform.
    
    While preparing the present manuscript we discovered another closely related paper introducing SuperstaQ software platform, which is dedicated to application-level quantum benchmarking~\cite{tomesh2022supermarq}. Meantime the focus of Arline platform is specifically benchmarking of quantum compilers, rather than a fully-fledged benchmarking of quantum computing architectures.

    \section{Platform Description}\label{sec:platform}
    Below we provide a quick overview  of Arline Benchmarks library functionality.  Arline Benchmark relies on a supplementary python package Arline Quantum as a backbone, which contains  a basic implementation of quantum gates, gate sets, quantum circuits and quantum hardware architectures. 
    \subsection{Architecture}
    Current implementation of Arline Benchmarks supports the following compilation/circuit optimization frameworks: Qiskit, Cirq, PyZX, Pytket (Python interface to Tket quantum compiler).
    \begin{figure}[H]
        \centering
        \includegraphics[scale=0.4]{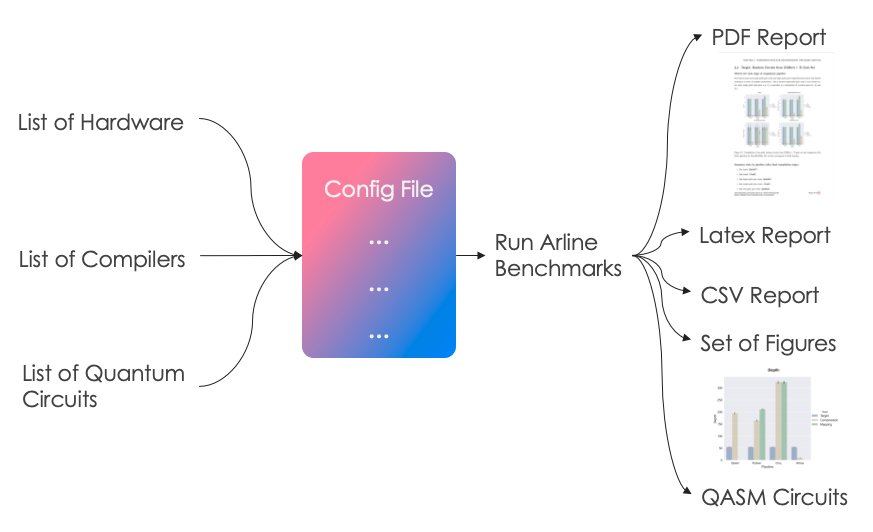}
        \caption{Workflow for generation of Arline Benchmarks reports.}
        \label{fig:report}
    \end{figure}
    Arline Benchmarks produces a PDF report file with detailed analytics of compilers’ performance for a user-defined configuration of benchmarking experiment, the schematic workflow is presented in Fig.~\ref{fig:report}.
    Each benchmarking experiment is configured by defining a list of hardware devices, list of target quantum circuits and list of compilation pipelines. In addition, Arline Benchmarks allows users to combine circuit compilation and optimization routines from different providers in a custom compilation pipeline to achieve the best performance.

    Arline Benchmarks runs result in the following  set of output files:
    \begin{itemize}
        \item Figures, charts and diagrams with relevant circuit metrics
        \item  Input/output QASM circuits corresponding to each compilation stage
        \item CSV report file with raw metrics for individual compilation stages
        \item LaTeX report file and final PDF report
    \end{itemize}    
    
\begin{figure}[H]
        \centering
        \includegraphics[scale=0.3]{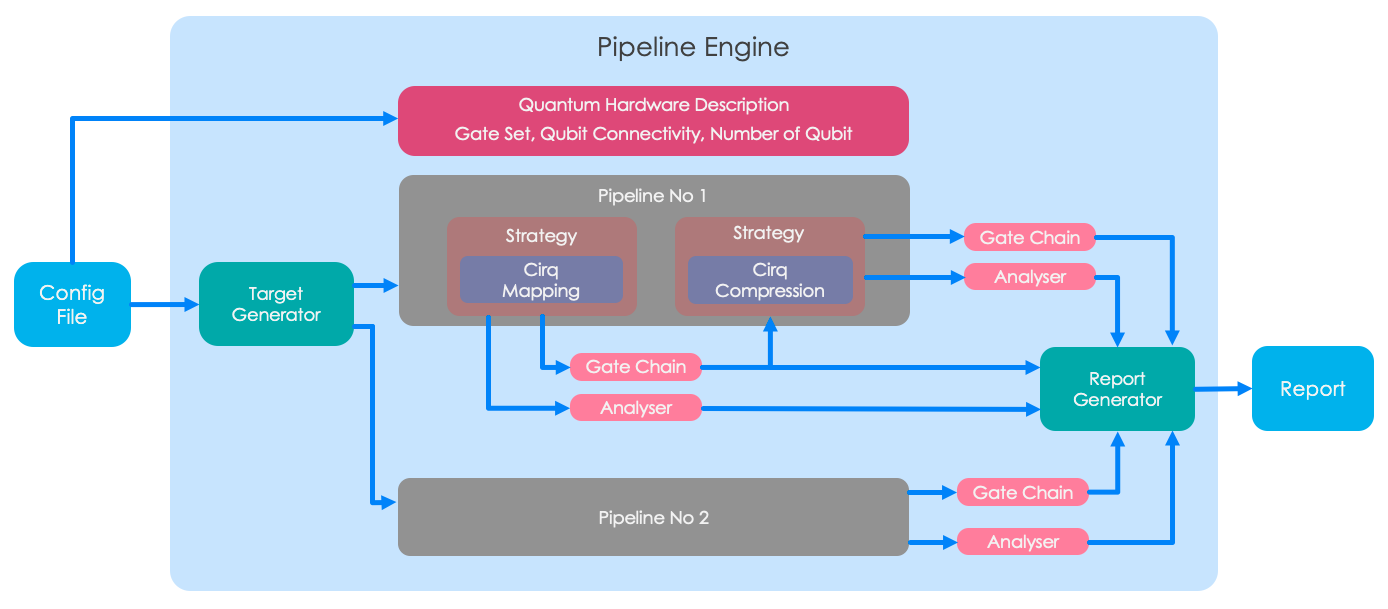}
        \caption{Arline Benchmarks software platform architecture. Arline's benchmarking engine sequentially executes compilation pipelines  for each target circuit and each hardware device. Relevant  metrics/meta-information for initial, final and intermediate quantum circuits produced during compilation are collected by analyser and sent to report generator, which produces a human-readable  benchmarking report in PDF format.}
        \label{fig:architecture}
    \end{figure}

    The architecture of Arline platform is illustrated in Fig.~\ref{fig:architecture}.
    User prepares configuration file in jsonnet format with the description of three main components of the benchmarking experiment: target circuits, quantum hardware and compilation pipeline. 
    \begin{enumerate}
        \item \textbf{Target} is a quantum circuit subject to compilation. Target circuits could correspond to  quantum algorithms loaded from OpenQASM files or  random quantum circuits generated from a specific gate set.
        \item \textbf{Hardware} configuration is specified by gate set, the number of qubits and qubit connectivity. Arline contains preconfigured gate sets for popular quantum hardware architectures (IonQ, Rigetti, IBM, Google).
        \item \textbf{Pipeline} is a sequence of compilation stages/subroutines (``strategies''). Arline's Strategy  is a wrapper class for subroutines incorporated from quantum compilation frameworks (e.g. Cirq Mapping, Qiskit Transpile etc). 
    \end{enumerate}
    
    \subsection{How to install Arline Benchmarks}
    In order to install Arline Benchmarks run in terminal
    \begin{lstlisting}
      pip install arline-benchmarks
    \end{lstlisting}
    Generation of PDF reports requires to install TeXLive distribution. 
    For further details on installation see   \href{https://github.com/ArlineQ/arline_benchmarks}{Arline's documentation}.
    In order to run your first benchmarking experiment execute
    \begin{lstlisting}{language=bash}
    bash run_and_plot.sh
    # run a simple benchmarking test and generate an automated report
    \end{lstlisting}
    
    The full description of the benchmarking experiment is stored in .jsonnet configuration file, which is an extension of .json data format.

    \subsection{Circuit Metrics}

    Quantum compilation consists of multiple stages such as translation of a quantum algorithm to gate instructions, qubit mapping and routing on connectivity-constrained architectures, circuit optimization, translation of gates to hardware native gate set, scheduling of quantum operations on real hardware.
    Circuit optimization/compression is a vital part of the compilation process. 

    A convenient way to measure circuit compression performance of a quantum compiler is to consider a ratio of input and output circuit metrics (e.g. gate count for a specific  gate type, circuit depth,   depth corresponding to gates from a specific type), which we call a \textbf{compression factor} or a \textbf{compression ratio}. Compression factor between two compilation stages stage$_{in}$ and stage$_{out}$ for a particular gate class $G$ is defined as
    
    \begin{equation}
        \textrm{ CF }(\textrm{gate count}, g\in G) = \frac{\textrm{gate count (stage$_{in}$})}{\textrm{gate count (stage$_{out}$})}, \qquad 
        \textrm{ CF (depth, $g\in G$)} = \frac{\textrm{gate depth (stage$_{in}$)}}{\textrm{gate depth (stage$_{out}$)}}.
        \label{eq:CF}
    \end{equation}
    
    In Eq. (\ref{eq:CF}) we assume a standard definition of a circuit depth and gate depth: circuit depth is the number of time slices corresponding to disjoint (parallelizable) operations in the circuit, and the gate depth refers to a circuit depth taking into account only  contribution of $G$-gates. In the context of NISQ devices, a relevant metric will be two-qubit count compression ratio or two-qubit gate depth. Compression ratio is the measure of relative reduction in the size of the circuit before and after circuit optimization. Compression factor smaller than unity $CF<1$ would correspond to an increase of the circuit size (gate count or depth) after performing circuit optimization. Such behaviour of circuit optimization algorithms is often undesirable. However, it is possible that an increase of one chosen metric (e.g. two-qubit gate count) will be accompanied with an increase of another metric (e.g. circuit depth) that makes a direct comparison of circuit optimization pipelines non-trivial. Typically, we will be interested in the compression ratio between the first stage corresponding to the input circuit and the final optimization stages.  The final output circuits for all compilation pipelines considered in the paper are translated (rebased) to the target hardware native gate set, that allows performing a fair comparison of compilation pipelines. Arline Benchmarks interface also supports the calculation of compression factor $CF$ between intermediate stages of a compilation pipeline, which could be helpful for a detailed analysis of separate compilation stages. 
    It is important to keep in mind that the single-qubit and two-qubit compression ratios are dependent on the input and output gate sets. 
    Single-qubit gate metrics are particularly sensitive to the choice of  input/output gate sets and should be interpreted with care.
    
    Ultimately, the most important performance metric is the proximity of the ideal and measured output bit-string probability distributions, which could be characterized by the fidelity of quantum operations on real hardware. The problem of estimation of a circuit fidelity is challenging and usually requires detailed knowledge of the noise model on  given quantum hardware, as well as simulating effects of noise on the operations on the quantum processor. 
    
    It is convenient to define a single quantity, a cost metric, that would incorporate information about compiled circuit (e.g. gate counts) as well as characteristics of the hardware, such as gate fidelities. We use the following definition of the \textbf{circuit cost function} $\mathcal{C}$:
    \begin{equation}
        \mathcal C = - D \log{K} - \sum_i \log{ F^{1q}_i} - \sum_j\log{F^{2q}_j},
        \label{eq:ibm_cost}
    \end{equation}
    where $\mathcal C$ - circuit cost, $D$ - circuit depth, $K$ - factor that penalises deep circuits,  $F^{1q}_i$ - fidelity of single-qubit gates, $F^{2q}_j$ - fidelity of two-qubit gates, the summation in (\ref{eq:ibm_cost}) is performed over all gates in the circuit. Our definition of the circuit cost function is equivalent to the minus logarithm of the cost function proposed by the IBM Quantum \cite{jurcevic2020demonstration}:
    \begin{equation}
        \mathcal C = - \log{C_{ibm}}, \quad \mathcal C_{ibm}=K^D\prod_i F^{(1q)}_i \prod_j F^{(2q)}_j
        \label{eq:C_ibm}
    \end{equation}
    
    Taking the logarithm in Eq. (\ref{eq:C_ibm}) helps to combat arithmetic underflow  when calculating a cost function of a large circuit since  the cost function in the multiplicative form shrinks to zero exponentially with an increase of the circuit size. The expression inside of log in Eq. (\ref{eq:ibm_cost}) could be interpreted as a total fidelity of the entire circuit assuming that the fidelity of a sequence of quantum gates can be factorized in the product of individual gates fidelities. We take the information about the fidelities of individual gates from published resources shared by vendors. An additional phenomenological parameter $K$ penalizes circuits with higher depth, that amounts to longer execution time on quantum hardware and higher decoherence rates. The phenomenological depth penalty parameter $K$ has a clear physical interpretation, however, it could be hard to estimate $K$ from the first principles based on the physical characteristics of the device, such as  $T_1$, $T_2$ coherence times,  gates execution times and the underlying noise model.
    We would like to note, that the proposed cost model (\ref{eq:C_ibm}) is still quite simplistic, although it is a convenient scalar metric to address the overall performance of a quantum algorithm on a given quantum hardware. More involved cost models should include the final measurement error, dependence of the depth penalty factor on  gate operations scheduling on the hardware, etc. In the future, handcrafted cost functions could be replaced  by neural-network-based approximators predicting hardware-specific noise levels for a given circuit~\cite{zlokapa2020deep}, although the predictive power of such models remains limited due to high variance  as well as such models will require a vast amount of training samples from real devices. 
    
    We present numerical values of average single-qubit, two-qubit gate fidelities and penalty parameter $K$ used in the present paper in Table \ref{table:cost_func_params}.  
    We choose the value of the penalty parameter $K$ from the range $F^{(1q)}<K<F^{(2q)}$.
    Parameter $K$ can be extracted empirically for a given on quantum hardware by collecting data on the fidelity of   
    random circuits with progressively increasing depth.
    More realistic circuit cost functions compared to Eq. (\ref{eq:C_ibm}) could be constructed by incorporating information about the scheduling of gate operations, accounting for cross-talk errors between individual gates or even considering non-Markovian noise models. 
    
    Note, that readout errors are not included in (\ref{eq:ibm_cost}), since the readout errors are identical in the original and optimized circuits, it is natural to disregard them in the definition of the circuit cost function. We define the circuit cost function improvement (ratio) as:
    \begin{equation}
        \textrm{Cost Improvement} = \frac{\textrm{Cost}(\textrm{stage$_{in}$})}{\textrm{Cost}(\textrm{stage$_{out}$})}.
        \label{eq:cost_improvement}
    \end{equation}

    We would like to emphasize that our benchmarking platform mostly focuses on NISQ devices, where the typical hardware native gate set contains continuous single-qubit rotations (e.g. $R_x(\phi)$, $R_z(\phi)$, $U_3(\theta, \phi, \lambda)$, etc.) and two-qubit entangling gates (e.g. $CX$, $CZ$, etc.). In contrast, for the future fault-tolerant devices with gates from Clifford+T gate set, the $T$-gate count is an appropriate cost metric of a quantum algorithm, and they require different benchmarking methodologies. In fault tolerant algorithms Clifford gates are used for quantum state operations and are cheap, whether the $T$-gates are typically used for magic state preparation/distillation and are expensive, so the cost function in fault-tolerant setting will require information about $T$-gate count and $T$-gate depth.
    Even though truly fault-tolerant regime implies unit perfect quantum operations, in the early fault-tolerant era quantum devices will still suffer from low amount of noise. Hence, it is likely that a hybrid cost function that incorporates information about both the $T$-gate metrics and the total fidelity will be necessary for benchmarking of nearly-fault-tolerant devices.
    
    \begin{table}[H]
        \caption{Gate fidelities and depth penalty factor $K$ used for the specification of the model cost function (\ref{eq:ibm_cost}).}
        \centering
        \begin{tabular}{|l|l|c|c|c|}
            \hline
            \textbf{Hardware Name} & \textbf{Native Gate Set} & $F^{1q}$ & $F^{2q}$ & $K$\\
            \hline 
            Mock IBM All2All & $CX$, $U_3(\theta, \phi, \lambda)$ & 0.9990 & 0.990 & 0.995 \\
            \hline 
            IBM Falcon 27q & $CX$, $U_3(\theta, \phi, \lambda)$ & 0.9996 & 0.990 & 0.995 \\
            \hline
            IBM Rueschlikon 16q & $CX$, $U_3(\theta, \phi, \lambda)$ & 0.9970 & 0.960 & 0.995 \\
            \hline
            Rigetti Aspen 16q & $CZ$, $R_z(\phi)$, $R_x(\pm k \pi/2)$, $k=1,2,3, \ldots$ & 0.9980 & 0.950 & 0.995\\
            \hline
            Google Sycamore 53q & $CZ$, $R_z(\phi)$, $R(\theta, \phi)$ & 0.9995 & 0.991 & 0.995 \\
            \hline
            IonQ 32q & $XX(\theta)$, $R_z(\phi)$, $R_x(\phi)$ & 0.9998 & 0.990 & 0.995\\
            \hline
        \end{tabular}
        \label{table:cost_func_params}
    \end{table}
    
    \subsection{Random Circuits Generation} \label{sec:random_circ_gen}
    
    Benchmarking of a quantum compilation stack on a real quantum hardware would require executing quantum algorithms that have a potential advantage over classical algorithms. However, most of promising quantum algorithms require prohibitively deep circuits, which are beyond the reach of current NISQ devices. Thus, random circuits is a natural choice for benchmarking of quantum compilers and quantum hardware from the practical standpoint.  For example, random circuits were used in Google's quantum supremacy experiment~\cite{arute2019quantum}. Random circuits are characterized by the gate set and the probabilities assigned to each gate type. We define a quantitative measure of the density of a particular gate class $G$:
    
    \begin{equation}
        \rho(g\in G) = \frac{gc(G)}{gc_{total}},
    \end{equation}
    where $gc$ is a gate count for the gates of class $G$ and $gc_{total}$ is the total gate count.
    
    All gates in the circuit are sampled independently from the uniform distribution with predefined probabilities $p_i(G)$. When sampling $CX_{ij}$ gates the locations $(i,j)$ of the control and target qubits are drawn from the uniform distribution: the control qubit is chosen from $i\in \{1, \ldots, N\}$, and the target qubit is chosen from the remaining $N-1$ qubits. We use two-qubit gate density $\rho(CX)$ to specify a particular class of random circuits and we set the probabilities of types of single-qubit gates to be equal. Continuous angles in single-qubit gates $U_{3}(\theta, \phi,\lambda)$ are sampled from the Haar distribution, and from the uniform distribution in case of the single angle rotation gates $R_z(\phi)$, $R_x(\phi)$, etc. As we will show in Sec. \ref{sec:random_circ_compression} the potential for the circuit compression strongly depends on the choice of the gate set and the two-qubit gate density of random circuits.

    \subsection{Overview of Circuit Optimization Algorithms}

    \begin{table}[H]
        \caption{Compilation subroutines used for benchmarking.}   
        \centering
        \begin{tabularx}{\textwidth}{|c|c|c|X|}
            \hline \textbf{Vendor} & \textbf{Subroutine} & \makecell{\textbf{Preserves} \\ \textbf{Connectivity} }  & \multicolumn{1}{c|}{\textbf{Description}} \\ \hline
            \multirow{4}{*}{\textbf{Qiskit}} & Transpile & $\checkmark$ & \noindent\parbox[c]{\hsize}{\strut The main Qiskit's compilation pass, that includes pre-processing of the input circuit (unroll of multi-qubit gates to single-qubit and two-qubit gates), performs mapping and routing to a given hardware connectivity, optimizes two-qubit subcircuits using KAK decomposition, removes gate-inverse pairs, removes diagonal gates before measurement gates.  Default settings for our benchmarking purposes correspond to heavy optimization (level 3) with SABRE routing algorithm. \strut}\\
            \cline{2-4}
            & Unroll & $\times$ & \noindent\parbox[c]{\hsize}{\strut Recursively expands three-qubit gates until the circuit only contains single-qubit or two-qubit gates. \strut}\\
            \cline{2-4}
            & CommutativeCancellation & $\checkmark$ & \noindent\parbox[c]{\hsize}{\strut Cancels the redundant (self-adjoint) gates through commutation relations. Utilizes the commutation relations between the following gates:  $H$, $X$, $Y$, $Z$, $CX$, $CY$, $CZ$. \strut}\\
            \cline{2-4}
            & KakBlocks & $\checkmark$ & \noindent\parbox[c]{\hsize}{\strut Searches and optimizes two-qubit subcircuits using KAK decomposition. \strut}\\
            \hline
            \multirow{6}{*}{\textbf{Pytket}} & Peephole & $\times$ & \noindent\parbox[c]{\hsize}{\strut Searches for non-optimal two-qubit subcircuits in the circuit graph and optimizes them in place by using KAK decomposition. Resulting circuit contains only $CX$, $U_1$, $U_2$ and $U_3$ gates. \strut}\\ 
            \cline{2-4}
            & PauliSimp & $\times$ & \noindent\parbox[c]{\hsize}{\strut Represents a circuit as a sequence of Pauli gadgets (aka phase gadgets) and a Clifford circuit, then resynthesises Pauili gadgets in groups of commuting terms. \strut}\\
            \cline{2-4}
            & SynthesiseIBM & $\checkmark$ & \noindent\parbox[c]{\hsize}{\strut Optimizes and converts all gates to $CX$, $U_1$, $U_2$ and $U_3$ gates. \strut}\\
            \cline{2-4}
            & DefaultMapping & $\checkmark$ & \noindent\parbox[c]{\hsize}{\strut Breaks circuit into time slices and inserts \textit{SWAP} gates to satisfy hardware connectivity constraints using a greedy-like algorithm with a finite depth look ahead. \strut}\\
            \cline{2-4}
            & PostRouting & $\checkmark$ & \noindent\parbox[c]{\hsize}{\strut Fast optimization pass, performing basic simplifications. If all multi-qubit gates are $CX$s, then this preserves their placement and orientation, so it is safe to perform after routing. \strut}\\
            \cline{2-4}
            & RemoveRedundancies & $\checkmark$ & \noindent\parbox[c]{\hsize}{\strut Removes gate-inverse pairs, merges rotations, removes identity rotations, and removes diagonal gates before measurement. \strut}\\
            \hline
            \multirow{2}{*}{\textbf{PyZX}} & FullReduce & $\times$ & \noindent\parbox[c]{\hsize}{\strut The main simplification routine of PyZX. It uses a combination of Clifford simplification and  gadgetization strategies. \strut}\\
            \cline{2-4}
            & FullOptimize & $\times$ & \noindent\parbox[c]{\hsize}{\strut Optimizes circuit using first basic commutation and cancellation rules, and then a dedicated phase polynomial optimization strategy. Supports only circuit from Clifford+T gate set.} \\
            \hline
            \multirow{7}{*}{\textbf{Cirq}} & OptimizeForXmon & $\checkmark$ & \noindent\parbox[c]{\hsize}{\strut Converts gates to $[CZ, U_3]$ gate set and optimizes circuit for Google transmon devices. \strut} \\
            \cline{2-4}
            & EjectPhasedPaulis & $\checkmark$ & \noindent\parbox[c]{\hsize}{\strut Commutes Phased Pauli gates $R_{x,y,z}$ to the end of the circuit and perform $CX$ optimization along the way. As the gates get pushed, they may absorb $R_z$ gates, get merged into measurements (as output bit flips), and add a global phase when crossing $CZ$ gates. \strut} \\
            \cline{2-4}
            & EjectZ & $\checkmark$ & \noindent\parbox[c]{\hsize}{\strut Commutes $R_z$ gates towards the end of the circuit. As the $R_z$ gates get pushed they may absorb other $R_z$ gates, get absorbed into measurement gates, cross $CZ$ gates, cross $R_x$ gates (by phasing them). \strut} \\
            \cline{2-4}
            & Merge1Q & $\checkmark$ & \noindent\parbox[c]{\hsize}{\strut Merges adjacent single-qubit gates. \strut} \\
            \cline{2-4}
            & MergeInteractions & $\checkmark$ & \noindent\parbox[c]{\hsize}{\strut Combines series of adjacent single and two-qubit gates operating on a pair of qubits and replaces two-qubit gates with CZ gates. \strut } \\
            \cline{2-4}
            & DropNegligible & $\checkmark$ & \noindent\parbox[c]{\hsize}{\strut Removes gates with small angles below tolerance value. \strut } \\
            \cline{2-4}
            & DropEmptyMoments & $\checkmark$ & \noindent\parbox[c]{\hsize}{\strut Removes empty time-slices (moments) from a circuit. \strut} \\
            \hline
            \makecell{\textbf{Arline} \\ \textbf{Benchmarks}} & Rebase & $\checkmark$ & 
            \noindent\parbox[c]{\hsize}{\strut Converts all gates to  hardware-native gate set.} \\
            \hline
        \end{tabularx}
        \label{tab:subroutines_full_list}\\
    \end{table}

    \begin{table}[H]
        \caption{List of QASM circuits types (quantum algorithms) used for benchmarking.}
        \centering
        \begin{tabularx}{\textwidth}{|c|c|c|X|}
            \hline
            \makecell{\textbf{Classes of} \\ \textbf{Quantum Algorithms}} & $n_{circ}$ & $N_{qubits}$ & \multicolumn{1}{c|}{\textbf{Description}} \\
            \hline
            Chemistry & 7 & 4-12 & \noindent\parbox[c]{\hsize}{\strut Unitary Coupled Cluster ansatz circuits for VQE algorithm (H$_2$, H$_2$O, LiH, NH, CH$_3$ molecules). \strut}\\
            \hline
            Quantum dynamics & 10 & 16 & \noindent\parbox[c]{\hsize}{\strut Trotterized evolution of a transverse field Ising model (TFIM), $H=\sum_{i} J_{i,i+1} Z_i Z_j + h_i X_i$, and TFIM with long range interactions, $H=\sum_{i\neq j} \frac{1}{|i-j|^\alpha} J_{ij} Z_i Z_j + h_i X_i$ for $N=20$ spins, number of Trotter steps $n_{Tr}=20$. Couplings $J_{i,i+1}$, $J_{ij}$, transverse magnetic field and the power-law exponent $\alpha$ were sampled from a uniform distribution: $J_{ij}\in [0, 1]$, $h_{i}\in [0, 1]$,  $\alpha\in[1,3]$.\strut}
            \\
            \hline
            Finance & 2 & 11 & \noindent\parbox[c]{\hsize}{\strut Amplitude amplification algorithm for option pricing estimation (European call/put option) \cite{woerner2019quantum, stamatopoulos2020option}. \strut}\\
            \hline
            Grover search & 4 & 9-15 & \noindent\parbox[c]{\hsize}{\strut Instances of Grover search algorithm  in Clifford+T gate set.\strut}\\
            \hline
        \end{tabularx}
        \label{tab:quantum_algos}
    \end{table}

    \begin{table}[H]
        \caption{List of compilation pipelines used for benchmarking and corresponding circuit optimization subroutines.}
        \centering
        \begin{threeparttable}
        \resizebox{\textwidth}{!}{\begin{tabular}{|c|c|c|c|}
            \hline
            \multirow{2}{*}{\textbf{Pipeline}} &
            \multicolumn{3}{c|}{ \textbf{Stages}}\\
            \cline{2-4}
            & \textbf{Compression Only} & \textbf{Mapping/Routing Only} & \textbf{Compression + Mapping/Routing}\\
            \hline 
            \makecell{QiskitPl} & \makecell{Transpile L3 (Qiskit) \\ Rebase (Arline)} & Transpile L0 (Qiskit) & \makecell{Transpile L3 (Qiskit) \\ Rebase (Arline)} \\
            \hline
            \makecell{PytketPl} & \makecell{Peephole (Pytket) \\ RemoveRedundancies (Pytket) \\  SynthesiseIbm (Pytket) \\ Rebase (Arline)} & \makecell{DefaultMapping (Pytket) \\ PostRouting (Pytket)} & \makecell{Peephole (Pytket) \\ DefaultMapping (Pytket) \\ PostRouting (Pytket) \\ RemoveRedundancies (Pytket) \\ SynthesiseIbm (Pytket) \\ Rebase (Arline)} \\
            \hline
            \makecell{PytketChemPl} & \makecell{PauliSimp (Pytket) \\ Peephole (Pytket) \\ SynthesiseIbm (Pytket) \\ Rebase (Arline)} & \makecell{--} & \makecell{PauliSimp (Pytket) \\ Peephole (Pytket) \\ DefaultMapping (Pytket) \\ SynthesiseIbm (Pytket) \\ Rebase (Arline)}\\
            \hline
            \makecell{CirqPl} & \makecell{OptimizeXmon (Cirq) \\ EjectZ (Cirq) \\ EjectPhasedPaulis (Cirq) \\ MergeInteractions (Cirq) \\ Merge1Q (Cirq) \\ DropNegligible (Cirq) \\ DropEmptyMoments (Cirq) \\ Rebase (Arline)} & \makecell{GreedyRouting (Cirq)} & \makecell{GreedyRouting (Cirq) or Transpile (Qiskit)\tnote{1} \hspace{0.1cm} \\ OptimizeXmon (Cirq) \\ EjectZ (Cirq) \\ EjectPhasedPaulis (Cirq) \\ MergeInteractions (Cirq) \\ Merge1Q (Cirq) \\ DropNegligible (Cirq) \\ DropEmptyMoments (Cirq) \\ Rebase (Arline)} \\
            \hline
            \makecell{PyZXPl} & \makecell{PyZXRebase (Arline) \\ FullReduce (PyZX) \\ Rebase (Arline)} & \makecell{--} & \makecell{PyZXRebase (Arline) \\ FullReduce (PyZX) \\ Transpile L0 (Qiskit) \\ Rebase (Arline)} \\
            \hline
        \end{tabular}}
        \begin{tablenotes}
            \item[1] \textit{GreedyRouting (Cirq)} timed out on some circuit instances and was replaced by \textit{Transpile (Qiskit)} subroutine.
        \end{tablenotes}
        \end{threeparttable}
        \label{table:pipelines}
    \end{table}

    \subsection{Quantum Hardware Architectures}

    \begin{figure}[H]
        \resizebox{\textwidth}{!}{\includegraphics{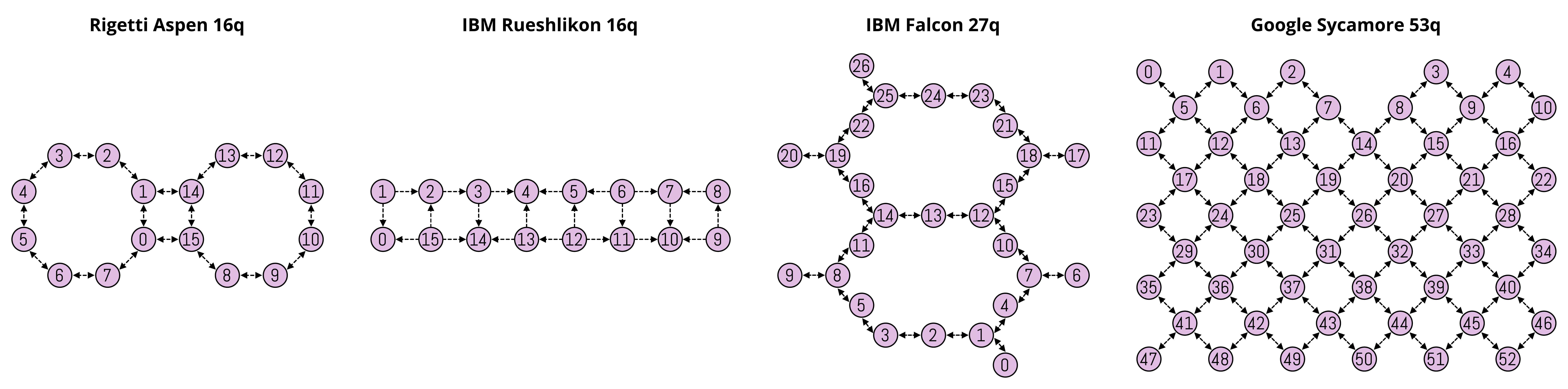}}
        \caption{Hardware connectivities used for benchmarking: Rigetti Aspen 16q, IBM Rueshlikon 16q, IBM Falcone 27q, Google Sycamore 53q.}
        \label{fig:hardware}
    \end{figure}
    
    \subsection{Circuit Equivalence Verification}
    
    In general, the verification of equivalence between the input $C_{in}$ and optimized $C_{out}$ circuits is QMA complete \cite{janzing2005non} and therefore is computationally hard (even for quantum computers!). The most straightforward (and not scalable) method to check if the two circuits are identical is to compute fidelity between unitaries of the input and output circuits: $F(U, V)=\frac{1}{2^N}|\mathrm{Tr} \left(U^\dag V\right)|$, where $U$ ($V$) are unitary $2^N\times 2^N$ matrices corresponding to input (output) circuits. If the unitaries $U$ and $V$ of the target and optimized circuits are equal up to a global phase, the fidelity reaches its maximal value, $F(U, V)=1$. However, some of the circuit optimization procedures do not preserve the target unitary operation and only preserve the output probability distribution $|\psi_{out}(\sigma_i)|^2$. An example of such optimization pass is \textit{RemoveDiagonalBeforeMeasurements} subroutine in \textit{Transpile (Qiskit)}, which removes diagonal gates in computational basis placed prior to terminal measurements. Thus, for circuit equivalence checking instead of fidelity between two unitaries we compute classical fidelity between two probability distributions:
    \begin{equation}
        \mathcal F_{cl} =\sum_{\sigma\in\{0,1\}^N} \sqrt{p_{in}(\sigma)p_{out}(\sigma)},
        \label{eq:F_cl}
    \end{equation}
    where $p_{in/out}(\sigma)=|\psi_{in/out}(\sigma)|^2$ are the  probability distributions of measured bitstrings $\{\sigma\}$, $\psi_{in/out}(\sigma)$ denote the $2^N$ dimensional statevector of target/optimized $N$-qubit circuits, respectively, summation in (\ref{eq:F_cl}) is performed over all bitstrings $\sigma$ of length $N$, that enumerate components of the statevector. In the case, if the input and output probability distributions of the measurement outcomes coincide, $p_{in}(\sigma)=p_{out}(\sigma)$, then the classical fidelity metric is equal to one, $\mathcal F_{cl}=1$. Arline Benchmark platform allows to compute fidelities $\mathcal F_{cl}$ for input circuit with a small number of qubits, $N \lesssim 15$, that could be used for debugging purposes. It is important to note, when compiling quantum circuit for hardware with restricted connectivity the qubit mapping/routing subroutines might introduce additional permutation relating logical and physical qubits. This permutation must be taken into account when computing the fidelity metric, Eq. (\ref{eq:F_cl}).
    
    A possible scalable approach for circuit equivalence checking is based on computational graph representation (DAG) of quantum circuits, e.g. Quantum Circuit Equivalence Checking (QCEC) project \cite{burgholzer2020advanced,burgholzer2020verifying}. Arline Benchmarks supports circuit equivalence checking via integration with QCEC module. However, it is important to note that for unstructured random circuits with continuous gates, such as Haar random circuits from $[CX, U_3]$ gate set will be still hard to verify. 
    In addition, future releases of Arline Benchmarks will include  Verified Optimizer for Quantum Circuits (VOQC) module, where equivalence checking is inbuild in the circuit optimization workflow  \cite{hietala2021verified}.

    \section{Warmup: two-qubit and three-qubit circuit optimization}
    \label{sec:kak+3q}
    \subsection{KAK Decomposition}

\begin{figure}[H]
        \centering
        \includegraphics[scale=0.5]{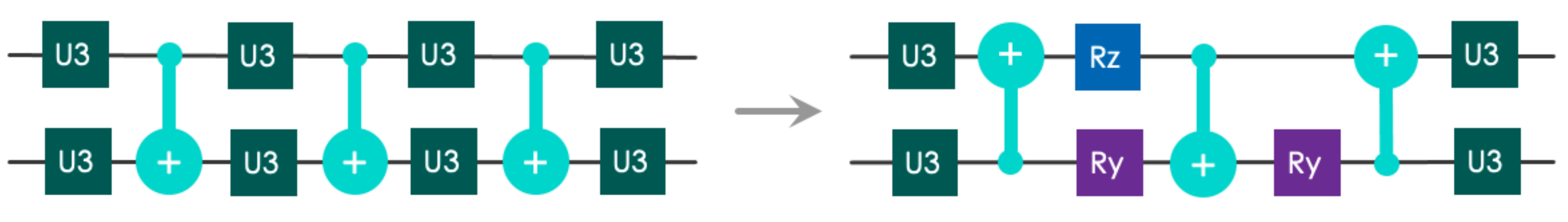}
        \caption{KAK (Cartan) decomposition of a generic two-qubit $SU(4)$ unitary.}
        \label{fig:kak}
    \end{figure}

    In this section, we consider toy benchmarking tests - compression of two-qubit and three-qubit circuits. Arbitrary two-qubit unitary $SU(4)$ can be represented as a circuit with no more than $gc(CX)=3$, $gc(U_3)=8$ gates and circuit depth $D = 7$. Such representation is known as Cartan decomposition or KAK decomposition and is proven to be $CX$-optimal \cite{dawson2004pra}. Each $U_3$ gate can be decomposed in a product of three single-axis rotations via Euler decomposition, e.g. $U_3 \sim R_z R_x R_z$. By exploiting the structure of KAK circuit it is possible to further reduce the number of single parameter gates (e.g. $R_z$ and $R_x$) from $gc(R_{x,z}) = 3\, gc(U_3)=24$ to $gc(R_{x,z}) = 15$ \cite{shende2003minimal}, see Fig. \ref{fig:kak}.

    Cartan decomposition is an important component of quantum compilers for the synthesis of arbitrary unitary gates for a particular hardware-native gate set. Besides that KAK decomposition is commonly used in compilers as a standard module for circuit compression/optimization of two-qubit subcircuits of large input circuits. For example, the Qiskit compilation passes \textit{Collect2qBlocks}, \textit{ConsolidateBlocks} and \textit{UnitarySynthesis}, which are the part of \textit{Transpile} combined optimization pass, perform the search of two-qubit blocks, postprocessing of the blocks and their resynthesis with the KAK decomposition. Similarly, subroutines \textit{Peephole} and \textit{SynthesisIBM} in Pytket compiler perform KAK resynthesis of two-qubit subcircuits.

\begin{figure}[H]
        \centering
        \resizebox{\textwidth}{!}{\includegraphics{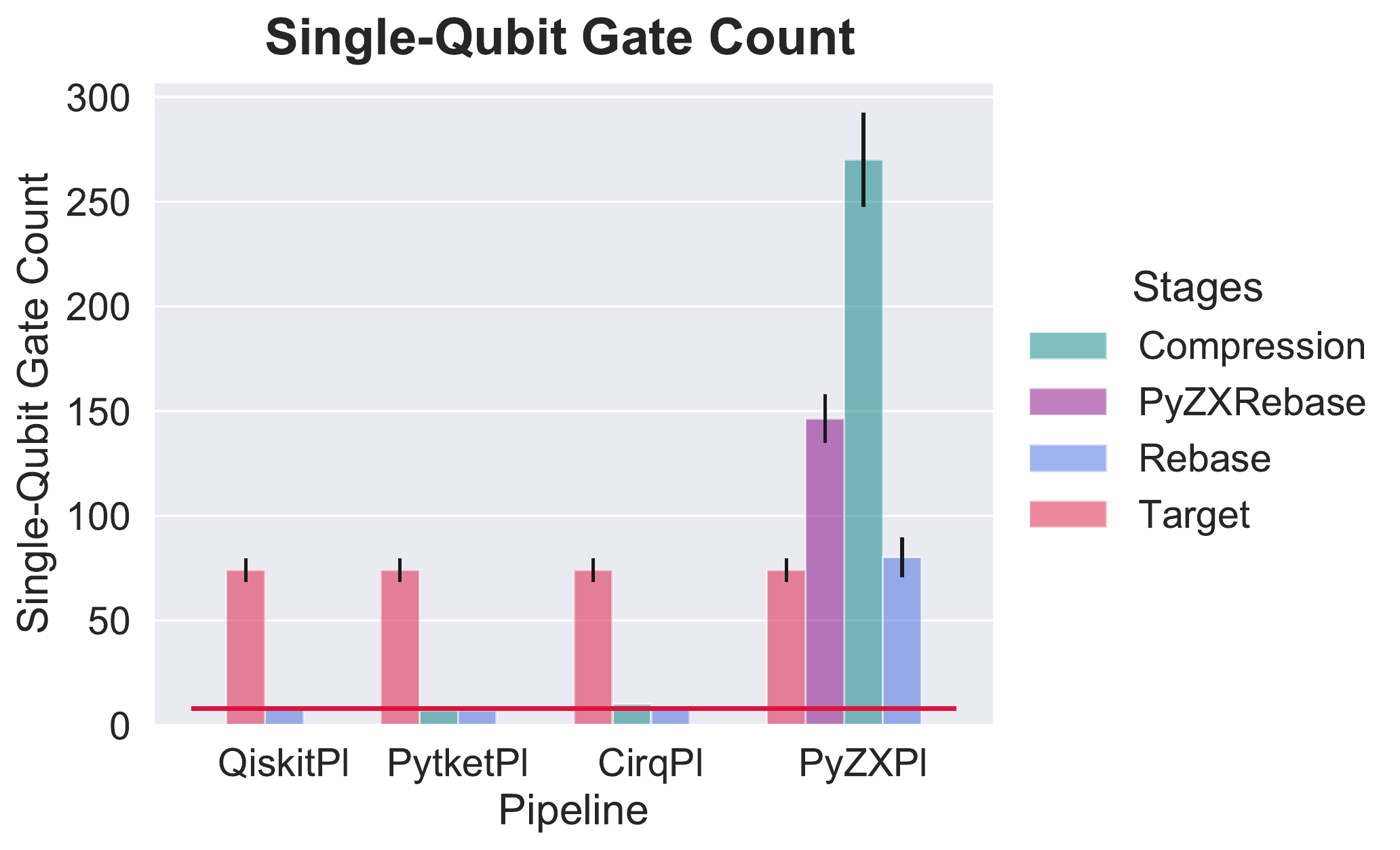}
        \includegraphics{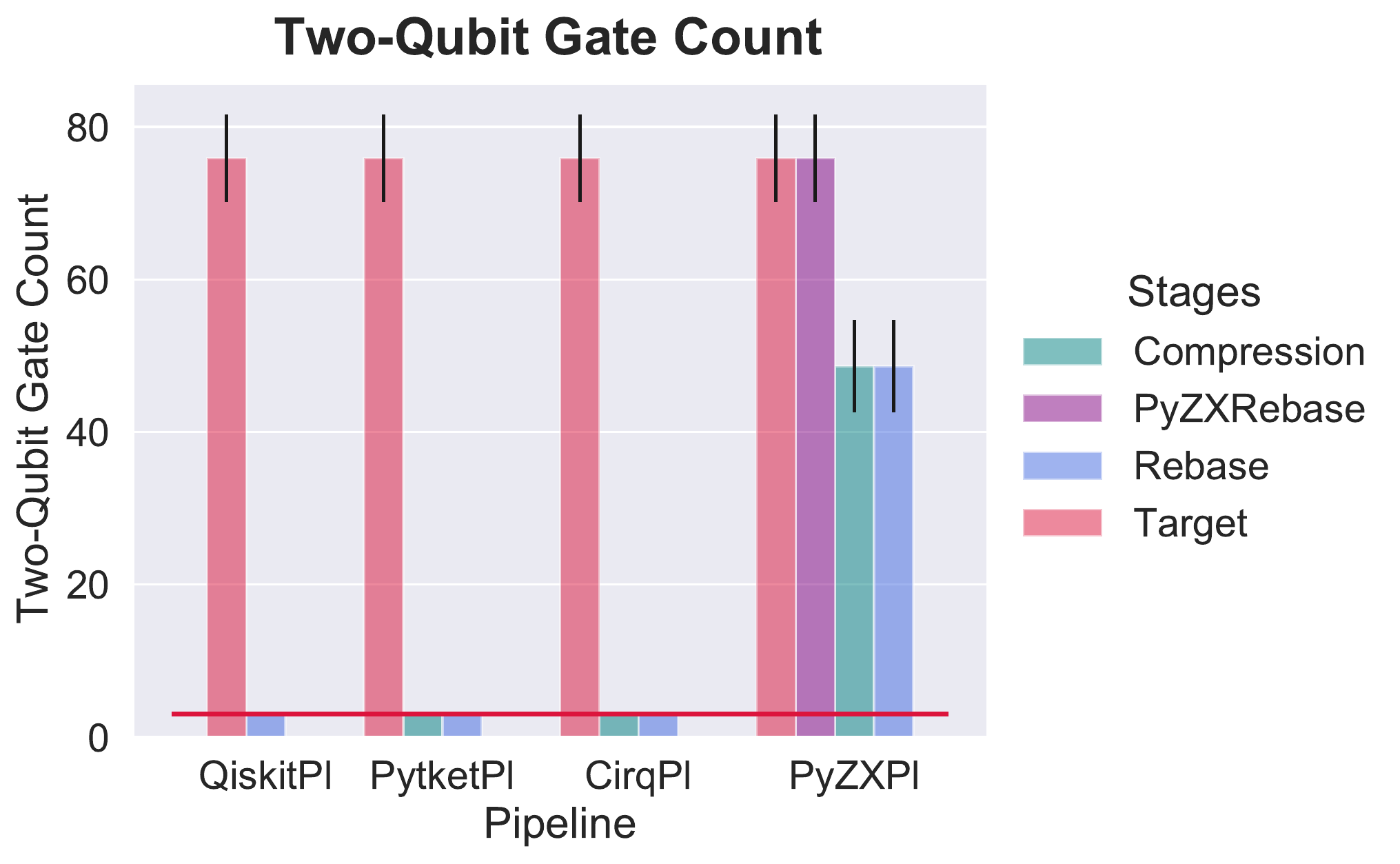}
        \includegraphics{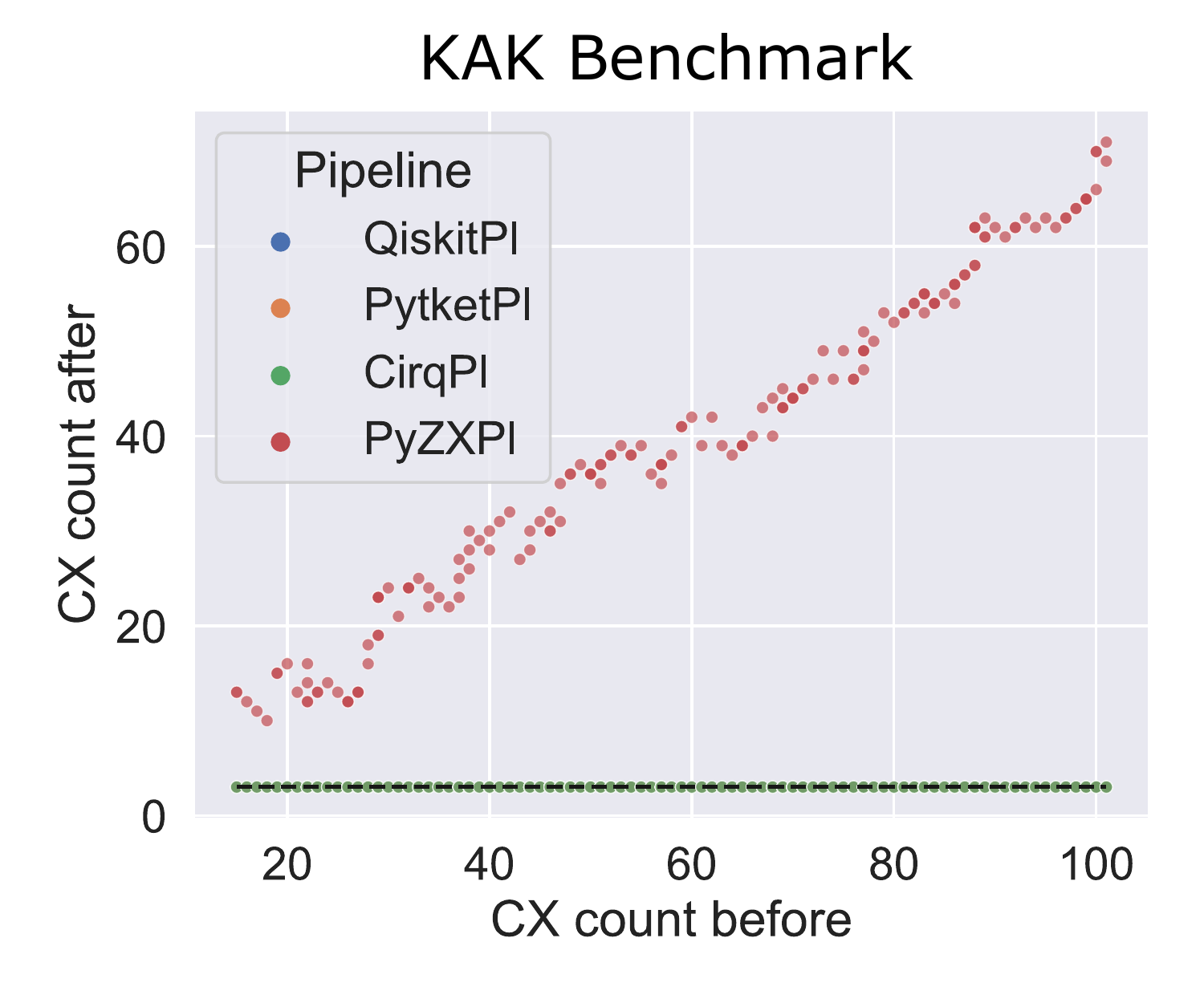}}
        \caption{KAK benchmark of two-qubit random circuits optimization, input circuit are sampled from $[CX, U_3]$ gate set. Metrics before/after compression: (a) $U_3$ gate count, (b) $CX$ gate count, (c) scaling of $CX$ gate count after circuit optimization  as a function of input circuit size. The initial circuit consisted of $g_{total}=150$ gates randomly selected from $[CX, U_3]$ gate set with gate densities: $\rho(U_3)=\rho(CX)=0.5$. The total number of sampled circuits $n_{circ}=30$.  Red horizontal line corresponds to the KAK optimal values for the single-qubit gate count $gc(U_3)=8$ and two-qubit gate count $gc(CX)=3$, see Fig. \ref{fig:kak}.
        Qiskit, Pytket and Cirq compilers achieved optimal performance after compression of two-qubit circuits coinciding with KAK decomposition.
        }
        \label{fig:2q_comparison}
    \end{figure}

    In Fig. \ref{fig:2q_comparison} we show the compression performance of quantum compilers for the KAK benchmark test for random two-qubit $[CX,U_3]$ circuits. We combine optimization subroutines in each compilation pipeline shown in Table \ref{table:pipelines} in a single optimization stage which we call \textit{Compression} and we transform the final optimized circuit to $[CX,U_3]$ gate set by \textit{Rebase (Arline)}. It is interesting to note that Qiskit, Pytket and Cirq were able to compress input circuits to the optimal ones with optimal $CX$ and $U_3$ gate counts. This shows that KAK decomposition is in-build in the compilation pipeline of Qiskit, Pytket and Cirq frameworks. 

    In contrast, the compilation pipeline based on PyZX module was not able to achieve optimal compression of two-qubit circuits and showed a large gap to optimality. Since PyZX framework supports only circuits with gates restricted to the following gate set [$H$, $S$, $S^\dag$, $T$, $T^\dag$, $CX$, $Z$, $R_x$, $R_z$, $CCX$, $CZ$] we perform rebase to the corresponding gate set prior to applying PyZX compression subroutines. We call the corresponding intermediate rebase stage as \textit{PyZXRebase}. The single-qubit gate count after PyZX \textit{Compression} stage was also far from optimal and was further reduced by \textit{Rebase} stage by fusing neighbouring single-qubit gates. From Fig. \ref{fig:2q_comparison}(c) we see that the output size of the circuit produced by PyZX framework grows linearly with the input size, that shows non-optimality of PyZX optimization subroutines for the two-qubit KAK benchmark. 
    
    \subsection{Three-Qubit Circuit Optimization}

    Another simple benchmarking test is a compression of three-qubit random circuits. The lower bound for the number of $CX$ necessary to implement a generic $SU(2^{N})$ unitary reads as \cite{shende2003minimal}: 
    \begin{equation}
    \label{eq:cx_bound}
        gc(CX)\geq \ceil{\frac{1}{4}(4^N - 3N - 1)}
    \end{equation}
    which gives a minimal $CX$ count for the worst case circuit on fully connected architecture, the corresponding bound is tight. The estimate for the single-qubit count could be derived from the simple dimension counting argument, that yields the minimum number of single-parameter gates   to be $gc(R_{x,z})\geq 2^{2N}-1$.

    In the limit of two-qubit circuits, $N=2$, this bound reduces to the prediction given by KAK decomposition with $g(CX)=3$ for arbitrary $SU(4)$ unitaries.
    Next, for three-qubit circuits, $N=3$, the theoretical optimal bound for the number of CNOT gates is $gc(CX) \geq 14$ \cite{shende2006synthesis} while the bound for the single-parameter one-qubit gates is $gc(R_{z,x})\geq 63$. We numerically verified the validity of this bound numerically for generic three-qubit unitaries by utilizing QFAST package \cite{younis2020qfast} for hierarchical quantum circuit synthesis. It is worth to note that bound (\ref{eq:cx_bound}) implies non-optimality of Quantum Shannon Decomposition scheme \cite{shende2006synthesis}, which requires $gc(CX)=24$ and the optimized version of Quantum Shannon Decomposition requires $gc(CX)=20$. The lower bound (\ref{eq:cx_bound}) grows exponentially with the number of qubits and becomes impractical for the purpose benchmarking of  quantum compilers for generic circuits with a large number of qubits, since the number of gates in typical circuits corresponding to quantum algorithms grows only polynomially with $N$.
    
\begin{figure}[H]
        \centering
        \resizebox{\textwidth}{!}{\includegraphics{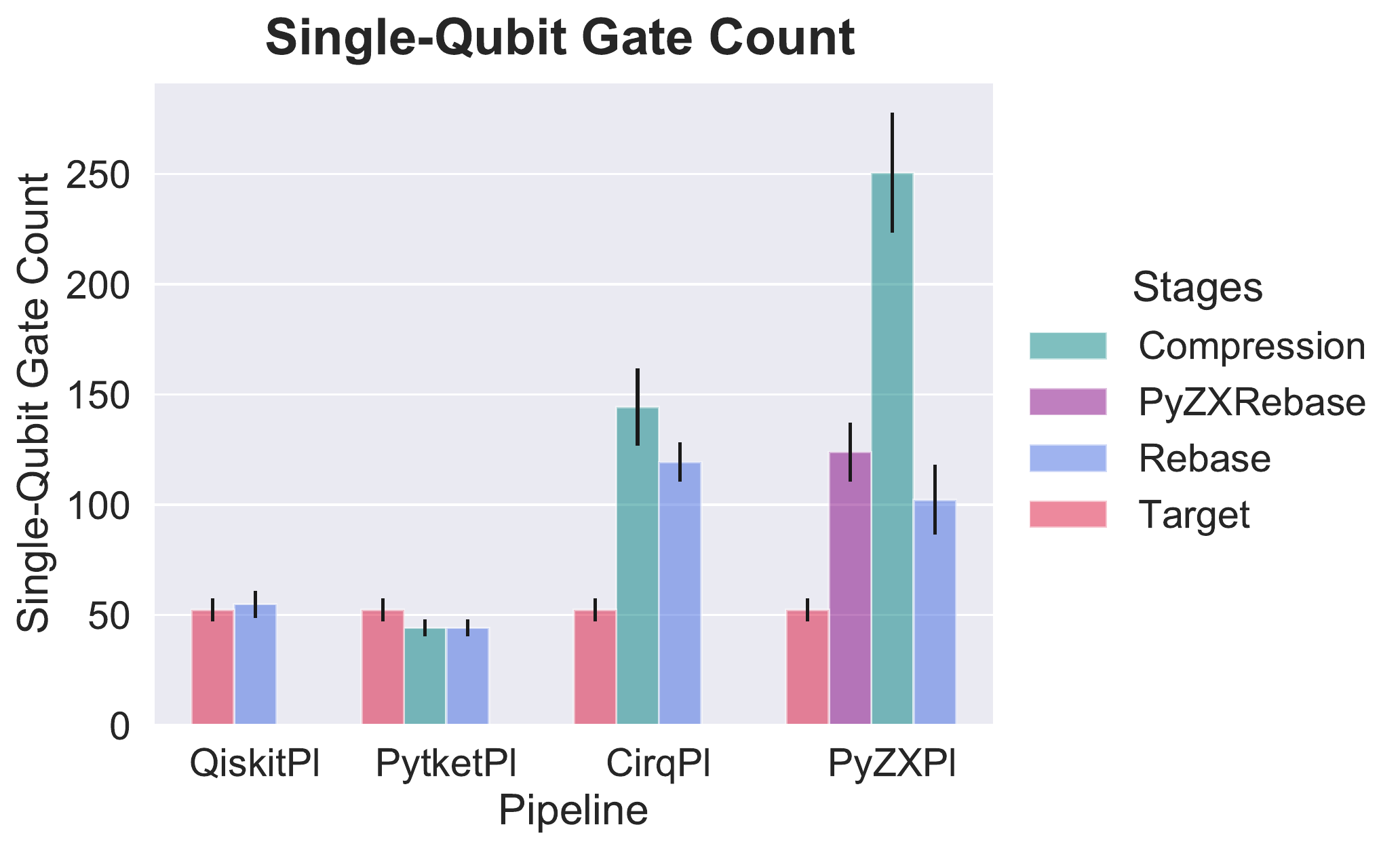}
        \includegraphics{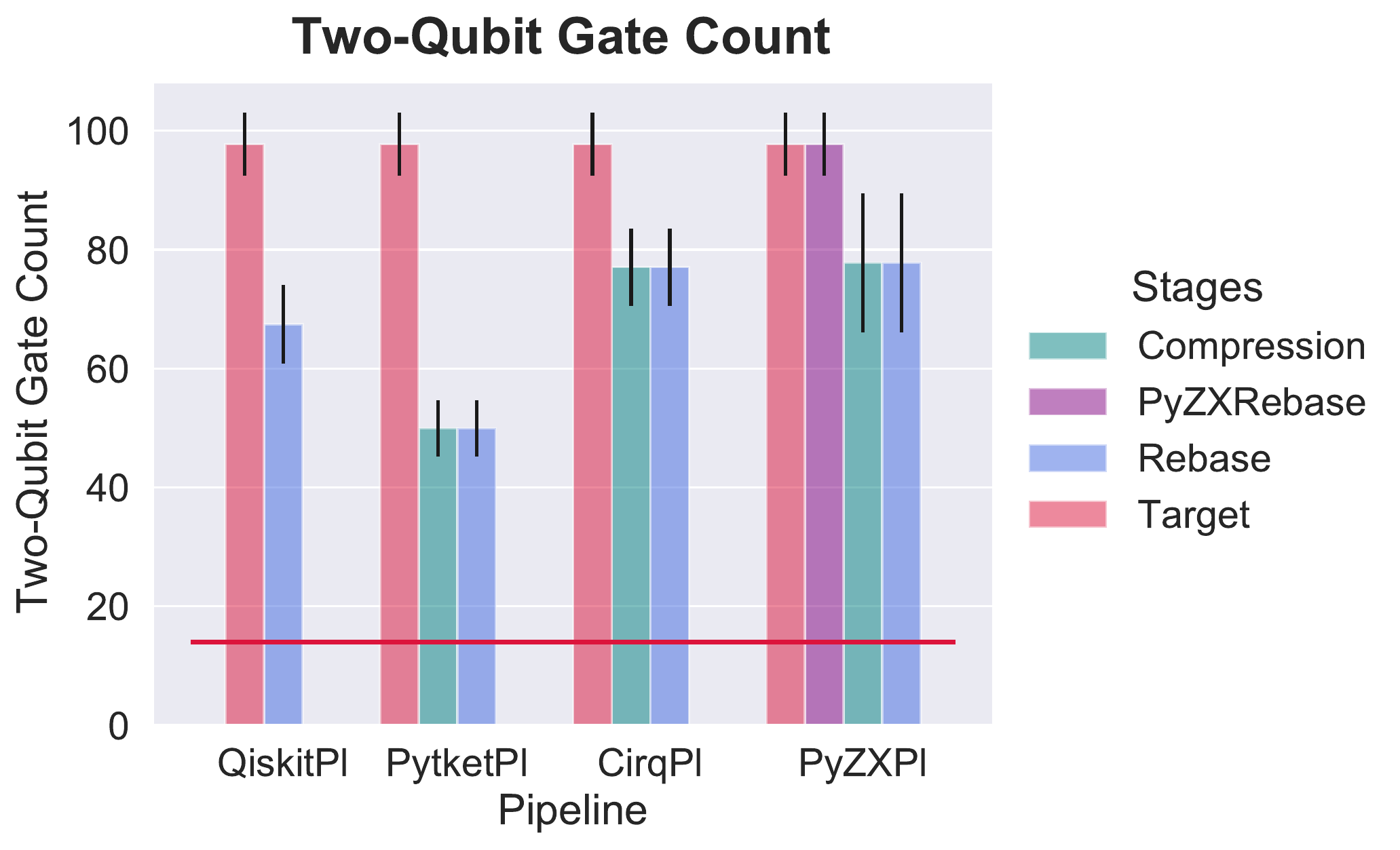}
        \includegraphics{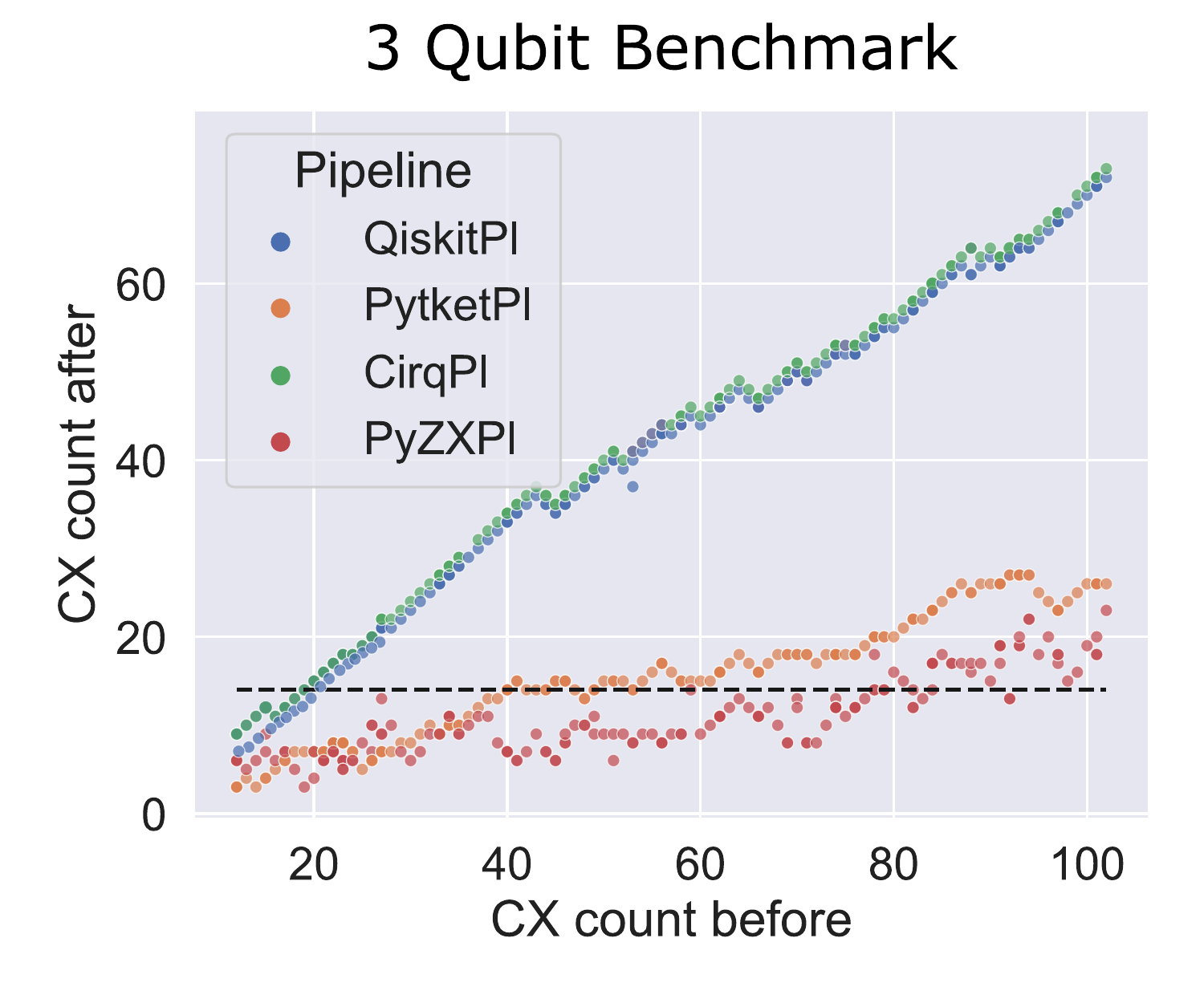}}
        \caption{Benchmarking of  three-qubit random circuits compression, circuits are sampled from $[CX, U_3]$  gate set. Metrics before/after compression: (a) single-qubit gate count ($U_3$), (b) two-qubit gate count ($CX$), (c) scaling of $CX$ gate count after circuit optimization  as a function of input circuit $CX$ count. The random circuits sampling procedure is the same as in Fig. \ref{fig:2q_comparison} and assumes all-to-all qubit connectivity graph. Circuits in (a, b) have a fixed size $g_{total}=150$, the number of random circuit samples and gate densities are the same as in Fig. \ref{fig:2q_comparison}. }
        \label{fig:3q_comparison}
    \end{figure}

    In Fig. \ref{fig:3q_comparison} we compare performance of quantum compilers for the task of compression of randomly generated three-qubit circuits. 
    Pytket achieved the best compression results for both $CX$ and single-qubit gate counts, although Pytket still was not able to reach the theoretical upper bound on the $CX$ count after compression (shown with the red horizontal line). Notably, Cirq demonstrates the worst performance and, both PyZX and Cirq increased the number of $U_3$ gates in the final circuit. In  Fig. \ref{fig:3q_comparison}(c) we show  the performance of compilers for input random circuits of varying depth. Note, that for the input circuits with low depth the output circuits could be implemented with less than $14$ CNOT gates, because the corresponding circuits are not generic. However, when increasing the depth of the input circuit the distribution of random unitaries approaches Haar distribution and the lower bound (\ref{eq:cx_bound}) becomes applicable.

    \section{Benchmarking of Routing/Mapping Subroutines}\label{sec:routing}
    
    Qubit routing and mapping subroutines are essential components of modern quantum compilers which are necessary for executing quantum circuits on hardware with restricted connectivity. Although sometimes both terms are used interchangeably, they describe different parts of the compilation process. ``Qubit routing'' refers to rewriting of multi-qubit gates in the quantum circuit to comply with the hardware connectivity graph. ``Qubit mapping'' subroutines map logical qubits $\{Q_{logic}\}$ specified in the quantum algorithm to physical qubits $\{Q_{phys}\}$ on real hardware. Qubit mapping procedure defines two permutations $P_{in}: \, \{Q_{logic}\}\to \{Q_{phys}\}$ and $P_{out}: \, \{Q_{phys}\}\to \{Q'_{logic}\}$ corresponding to mapping of logical qubits to physical in the beginning of the circuit and mapping of physical qubits back to logical qubits at the end of the circuit after terminal measurement gates. Permutations $P_{in}$ and $P_{out}$ are not necessarily inverse of each other since $P_{out}$ could be absorbed into reindexing of classical bits that store results of measurement gates.
    
    First, we perform benchmarking of mapping/routing  subroutines of Qiskit, Pytket and Cirq frameworks for randomly generated qubit connectivity graphs. In order to study routing/mapping subroutines performance for different sparsity of the qubit connectivity graph, we consider random $k$-regular graphs with varying node degree $k$. As a relevant metric, characterizing overhead introduced by the routing/mapping subroutine we consider the following ratios
    \begin{equation}
        \textrm{Gate Count Overhead} = \frac{\textrm{gate count (stage$_{out}$)} }{\textrm{gate count (stage$_{in}$)}}, \qquad
        \textrm{Depth Overhead} = \frac{\textrm{depth (stage$_{out}$)} }{\textrm{depth (stage$_{in}$)}},
    \end{equation}
    where stage$_{in}$ refers to the input circuit and stage$_{in}$ corresponds to the output circuit after routing/mapping pass. The connectivity graph could be either directed or undirected, depending on the hardware constraints.

    
    In practice, quantum programmers are often interested in average case routing performance instead of the worst case performance. Moreover, although the \textit{SWAP} overhead is a useful metric to characterize routing overhead on a given hardware graph, in reality, quantum programmers are mostly interested in the $CX$ count overhead. Theoretical tools for understanding and characterizing $CX$ overhead of routing algorithms on various hardware graphs are quite scarce. 

\begin{figure}[H]
        \centering
        \resizebox{\textwidth}{!}{\includegraphics[scale=0.9]{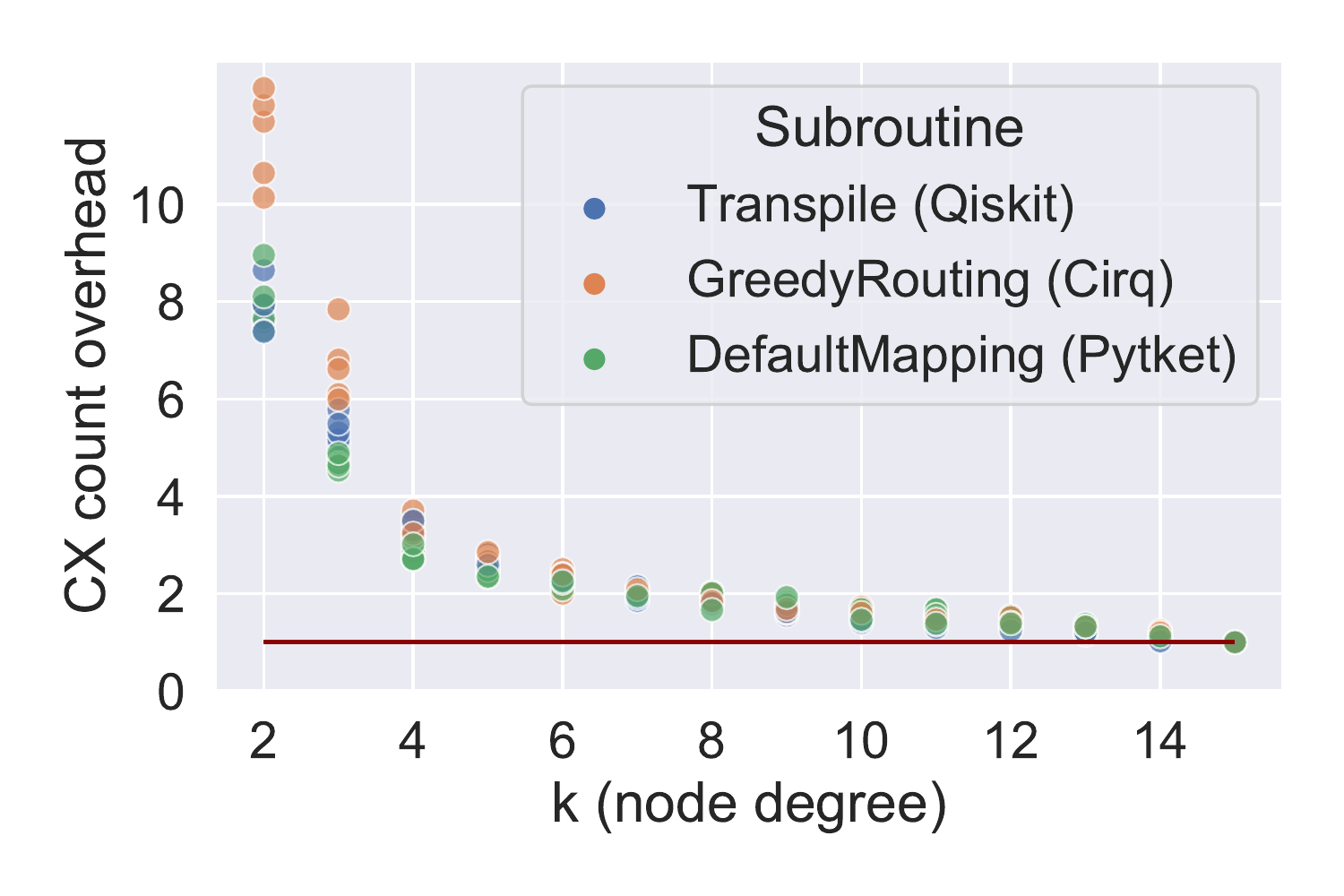}
        \includegraphics[scale=0.9]{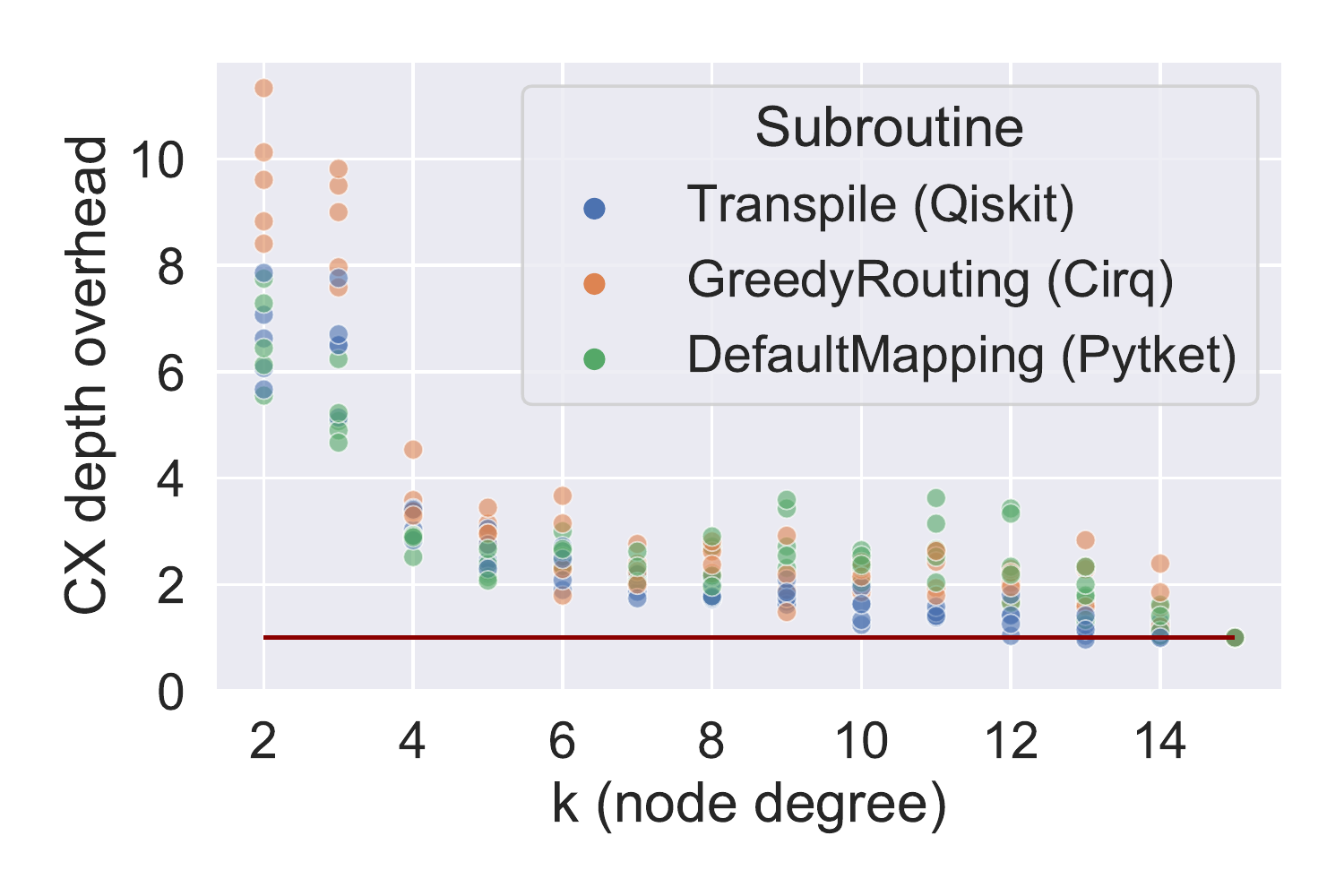}}
        \caption{Benchmarking of routing/mapping subroutines on $k$-regular random  graphs with $N=16$ nodes. (a) $CX$ count overhead (b) $CX$ depth overhead introduced by the routing/mapping subroutines. The red horizontal line shows a baseline with no multiplicative overhead ($\textrm{overhead}=1$). For each value of $k$ we generate a single instance of a random $k$-regular graph. The target circuits are random circuit from $[CX, U_3]$ gate set, with the total gate count $g_{total}=300$, each point on the plot represent a random circuit instance.}
        \label{fig:cx_k-regul_random}
    \end{figure}

    We test the performance of three routing/mapping subroutines: \textit{Transpile (Qiskit)} with \textit{SWAP}-based bidirectional heuristic search algorithm (SABRE) \cite{li2019tackling},  \textit{DefaultMapping (Pytket)} subroutine that performs dynamical remapping of qubits in each time slice \cite{cowtan2019qubit} and \textit{GreedyRouting (Cirq)} swap-network based greedy algorithm \cite{cirq}. In Fig. \ref{fig:cx_k-regul_random} we show the dependence of $CX$ count and $CX$ depth overhead on the graph node degree $k$ for randomly generated undirected $k$-regular graphs. The limiting case of node degree $k=2$ correspond to a ring connectivity that results in the maximal overhead (linear $CX$ count and $CX$ depth overhead for the worst case instances), and $k=N-1$ corresponds to the all-to-all qubit connectivity with no overhead. Overall, the performance of Qiskit and Pytket routing/mapping subroutines is very similar for a wide range of values of $k$ according to $CX$ count overhead metric. Both algorithms show somewhat different performance according to $CX$ depth overhead, where \textit{Transpile (Qiskit)}  outperforms \textit{DefaultMapping (Pytket)} for densely connected graphs with node degrees $k \gtrsim 8$.

\begin{figure}[H]
        \centering
        \resizebox{\textwidth}{!}{\includegraphics{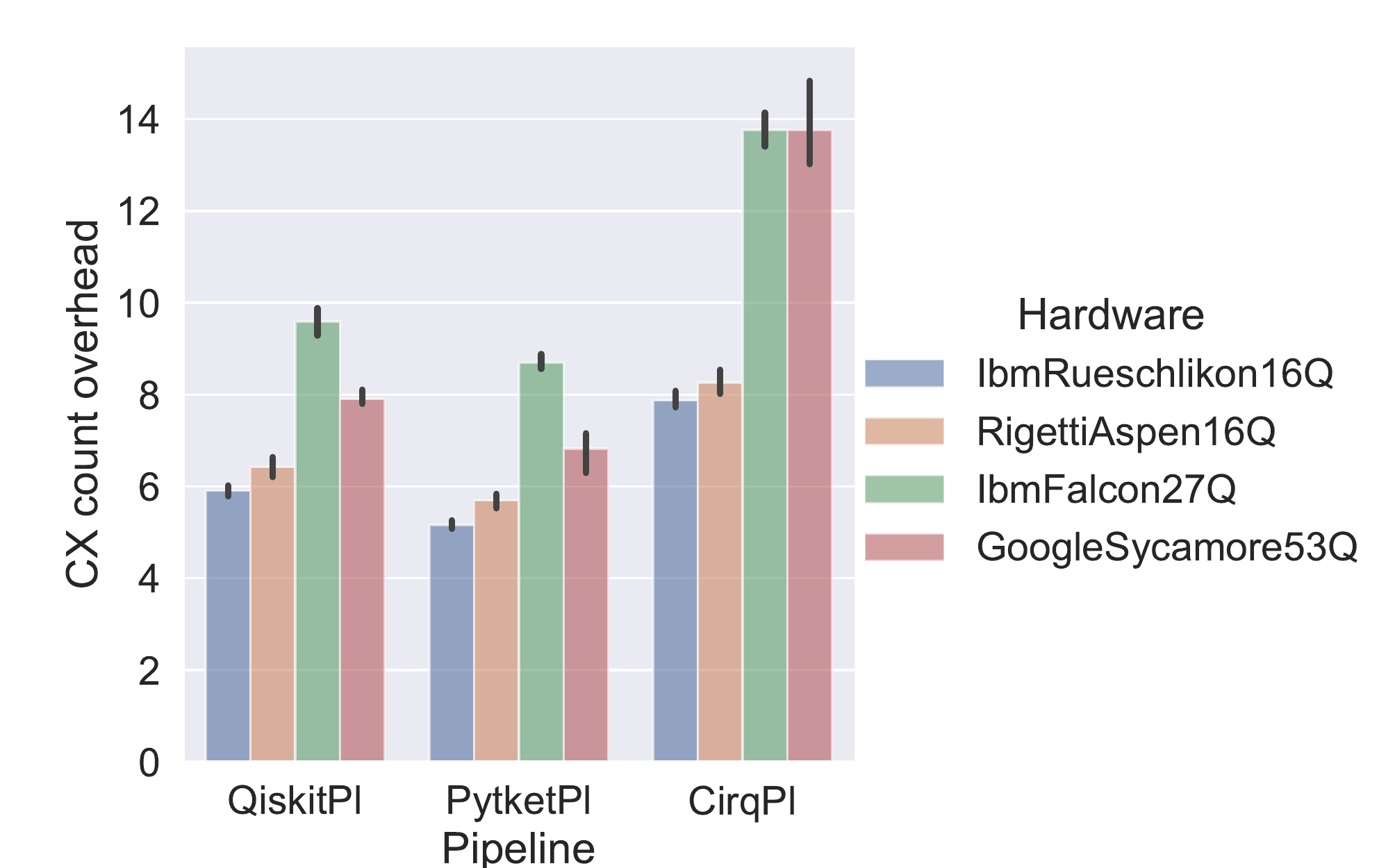}
        \includegraphics{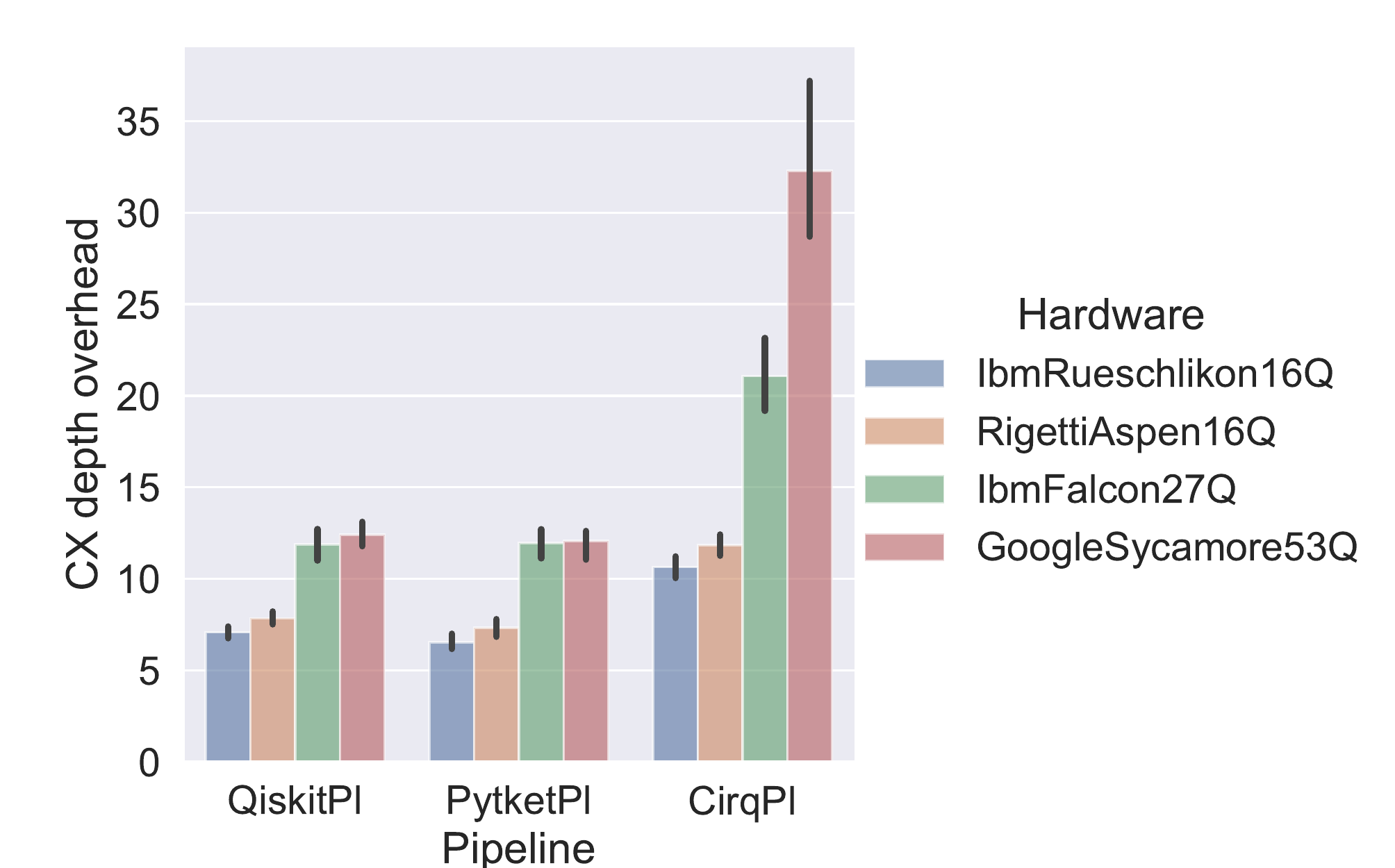}}
        \caption{Comparison of overheads introduced by routing/mapping compilation subroutines for four hardware architectures: Rueshlikon 16q, Rigetti Aspen 16q, IBM Falcon 27q, Google Sycamore 53q, see Fig. \ref{fig:architecture}. The target circuits are generated using the same parameters as in Fig. \ref{fig:cx_k-regul_random}.}
         \label{fig:cx_hardware}
    \end{figure}

    Besides routing on random graphs, it is interesting to benchmark routing subroutines for coupling graphs corresponding to real devices. The resulting overheads for four popular hardware  architectures are shown in Fig. \ref{fig:cx_hardware}. The relative performance of Qiskit, Pytket and Cirq frameworks strongly correlates with the corresponding results for routing on random $k$-regular graphs (Fig. \ref{fig:cx_k-regul_random}): \textit{Transpile (Qiskit)} and \textit{DefaultMapping (Pytket)} demonstrate approximately matching performance with a slight edge by Pytket. \textit{GreedyRouting (Cirq)} showed largest overheads for all target circuit types and hardware architectures.
    
    As a side note, we would like to point out that the $CX$ (gate count, depth) overhead metric is insensitive whether the coupling  graph is directed or undirected. This is due to the standard identity $CX_{ij}=(H_i \otimes H_j) CX_{ji} (H_i \otimes H_j)$ which allows reversal of control-target polarity of $CX$ gate by sandwiching it with Hadamard gates $H$.
    
    \section{Circuit Compression}\label{sec:compression}
    
    \subsection{Random Circuits} \label{sec:random_circ_compression}
    
    In this section, we compare compression performance of Qiskit, Pytket, Cirq and PyZX frameworks using random circuits as targets. The methodology of random circuit generation is described in Sec. \ref{sec:random_circ_gen}. In order to separate effects associated with routing/mapping,  we will assume all-to-all hardware connectivity, that allows us to remove corresponding subroutines from compilation pipeline, see ``Compression only'' column in Table \ref{table:pipelines}. For circuit cost function evaluation, we use parameters corresponding to Mock IBM All2All hardware in Table \ref{table:cost_func_params}. We will consider random target circuits of three types:
    
    \begin{itemize}
        \item[(i)] Random $[CX, U_3]$ circuits with a low  density of $CX$ gates, $\rho(CX)=0.1$;
        \item[(ii)] Random $[CX, U_3]$ circuits with a high  density of $CX$ gates, $\rho(CX)=0.9$;
        \item[(iii)] Random phase polynomial $[CX, R_z]$ circuits with a high $CX$ density, $\rho(CX)=0.9$.
    \end{itemize}
    
\begin{figure}[H]
        \includegraphics[scale=0.4]{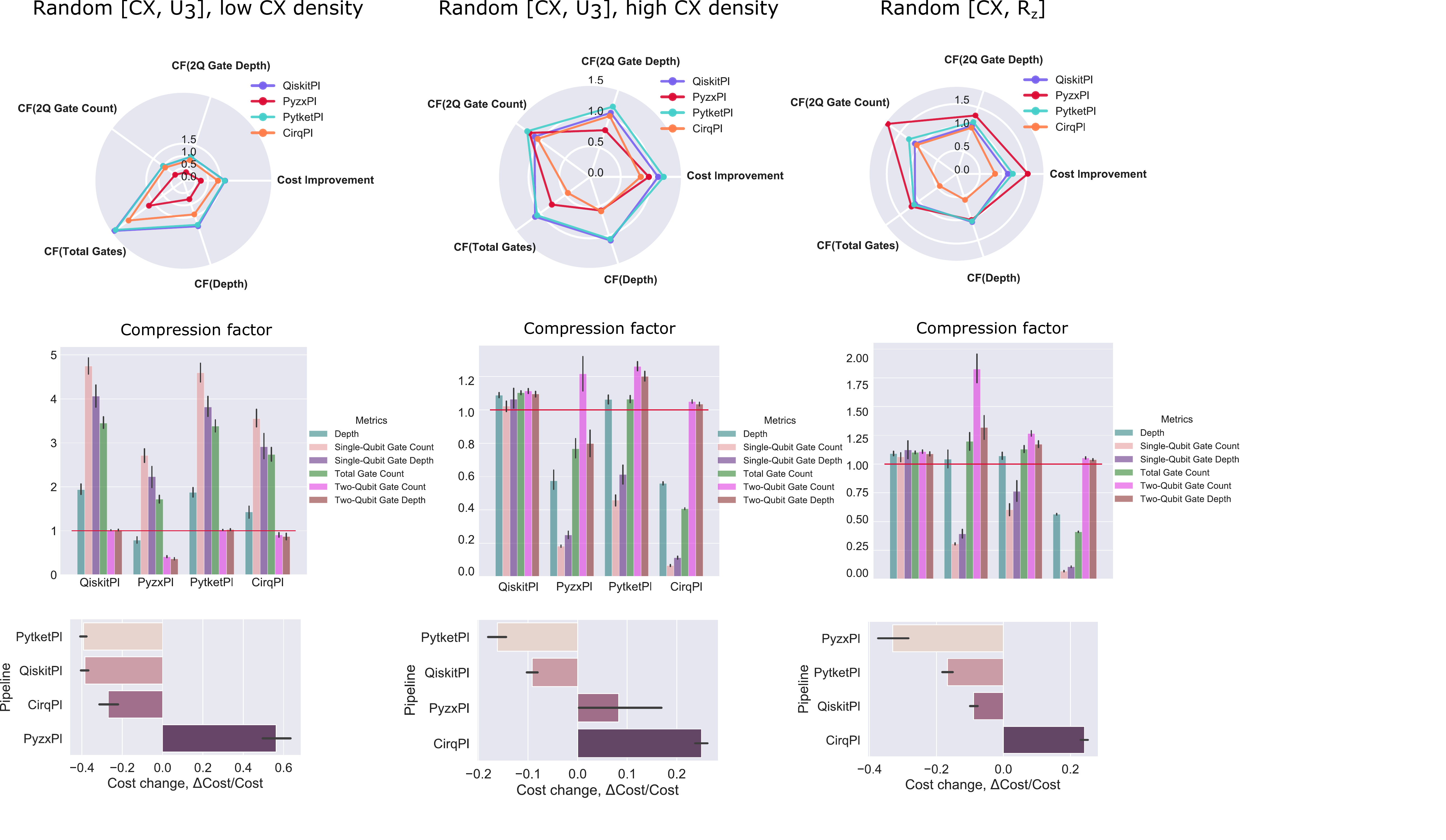}
        \caption{Compression performance for three classes of random circuits: (i), (ii), (iii) correspond to the first, second and third columns. Top row: Aggregate statistics for average compression metrics, radar-plot representation. The best compression performance corresponds to the largest polygon area on the radar plots. Middle row:  Compression factor $CF$ computed between the initial and the final (rebase) stages. Bottom row: Relative change of the circuit cost function $\Delta \mathcal{C}/\mathcal{C} = (\mathcal{C}_{out}-\mathcal{C}_{in})/\mathcal{C}_{in}$, Eq. \ref{eq:ibm_cost}. Parameters of hardware cost function correspond to Mock IBM All2All 10q device, see Table \ref{table:cost_func_params}. The compression ratios are averaged over $n_{circ}=20$ circuit instances with total $gc_{total}=300$ gates in each circuit.}
        \label{fig:random_combo}
    \end{figure}

    In contrast to $[CX, U_3]$ gate set which is universal, the gate set $[CX, R_z]$ is not universal. However, $[CX, R_z]$ circuits have interesting theoretical properties: such circuits can be concisely represented in terms of phase polynomials \cite{amy2013meet, nam2018automated, meijer2020architecture} that allows efficient circuit compression \cite{amy2018controlled}. In Fig. \ref{fig:random_combo} we show an aggregate compression performance results using radar-plot representation. Analysing Fig. \ref{fig:random_combo} we arrived to the following observations:
    
    \begin{itemize}
        \item[(a)]  For random $[CX, U_3]$ circuits with low $CX$ density there is a significant reduction of the single-qubit gate count and total gate count ($CF\sim  2-4$), but almost no change in the $CX$ count except of PyZX pipeline, where the final $CX$ count increased by a factor of $\sim 2$ after optimization, $CF(CX\: \textrm{count})\sim 0.5$. Such behaviour is quite expected, since in circuits with a low density of $CX$ gates and high density of single-qubit gates it is more likely to find single-qubit gate sequences that could be further optimized, and it is unlikely to find two-qubit gate sequences amenable to optimization. Qiskit and Pytket pipelines showed the best overall performance.
        \item[(b)] For random $[CX, U_3]$ circuits with high $CX$ density the reduction of $CX$ gates is about $10-20\%$. Qiskit and Pytket showed the best performance.
        \item[(c)] For random phase polynomial $[CX, R_z]$ circuits, PyZX outperforms competitors by a large margin in terms of $CX$ count reduction, although by the expense of increasing of single-qubit gate count. This behaviour is expected, since in \textit{FullReduce (PyZX)}  subroutine  relies on phase polynomial representation quantum circuits by exploiting identity relations between Pauli gadgets. Pauli gadgets are multi-qubit unitaries constructed by exponentiation of a Pauli string acting on a subset of qubits, e.g. $P(\theta, \vec\mu ) = \exp{\left(i\theta\, \sigma_1^{\mu_1} \otimes \sigma_2^{\mu_2} \otimes \ldots \sigma_k^{\mu_k}\right)}$, where $\sigma_i^{\mu_i\in [x,y,z]}$ are Pauli matrices. 
    \end{itemize}    

    As for the most of our benchmarking tests, it is hard to identify an absolute winner, because the compression performance strongly depends on the circuit target type. Thus the relative ranking of compilation pipelines is meaningful only for quantum circuits from a specific class.
    
\begin{figure}[H]
        \resizebox{\textwidth}{!}{\includegraphics{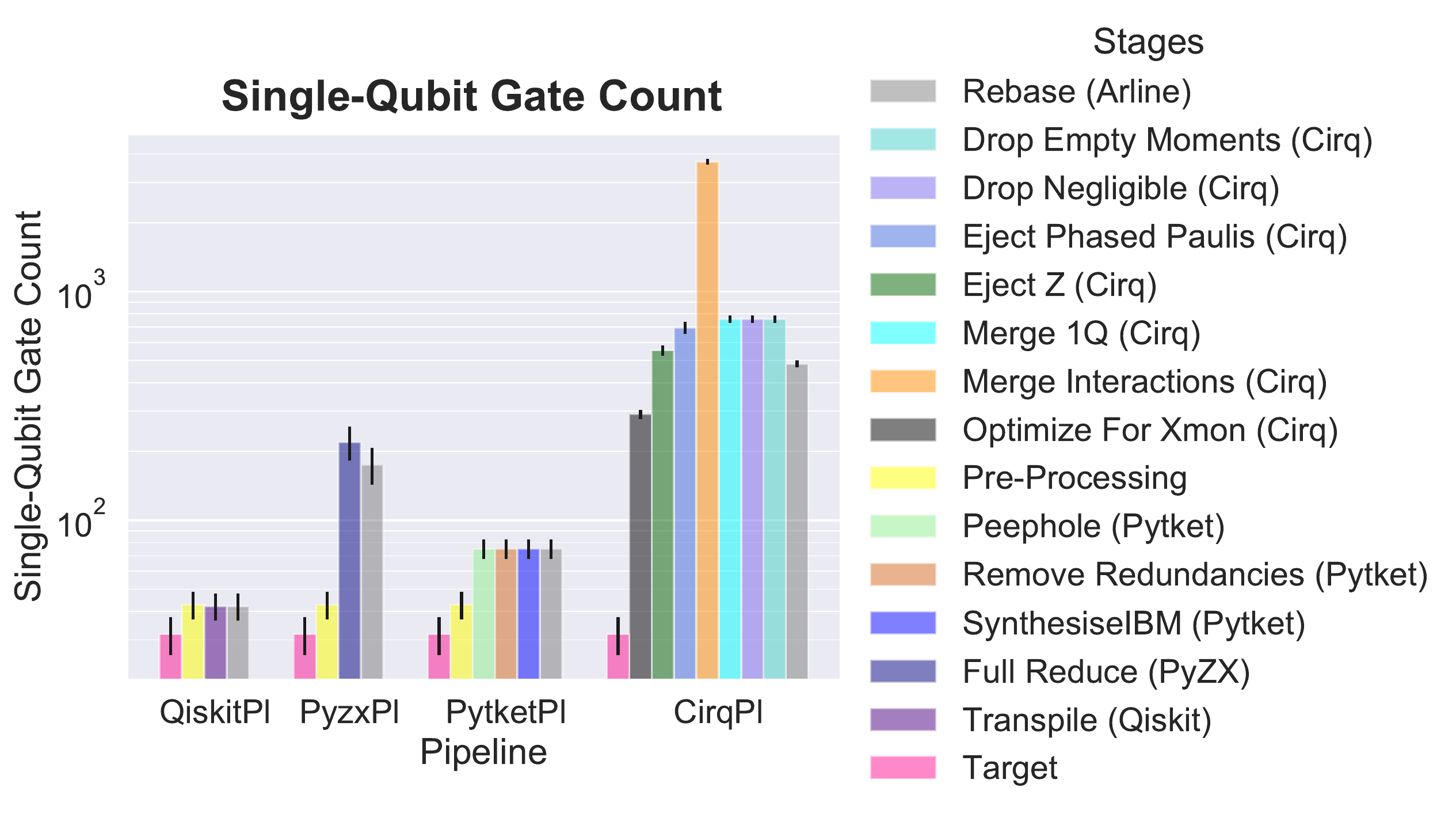}
        \includegraphics{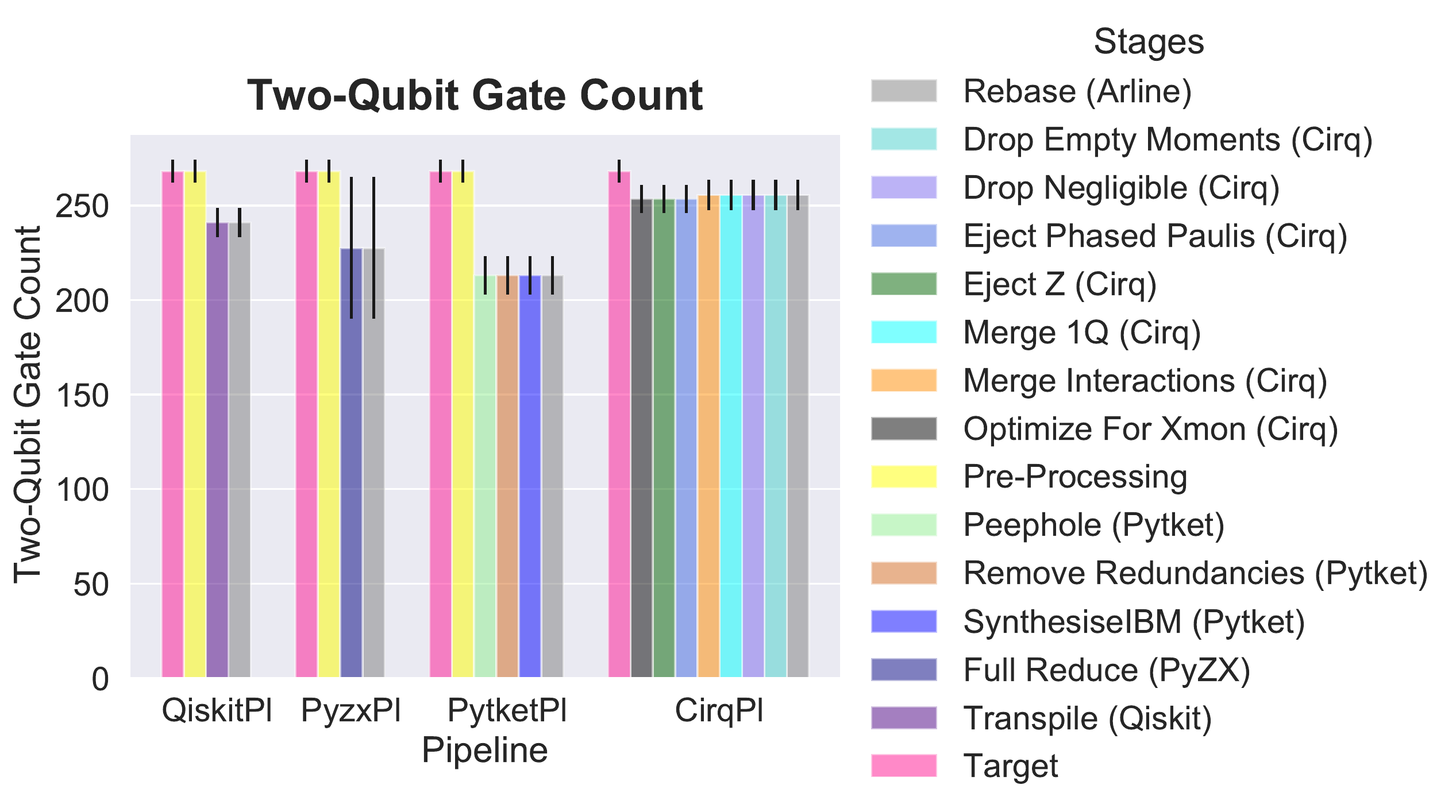}
        \includegraphics{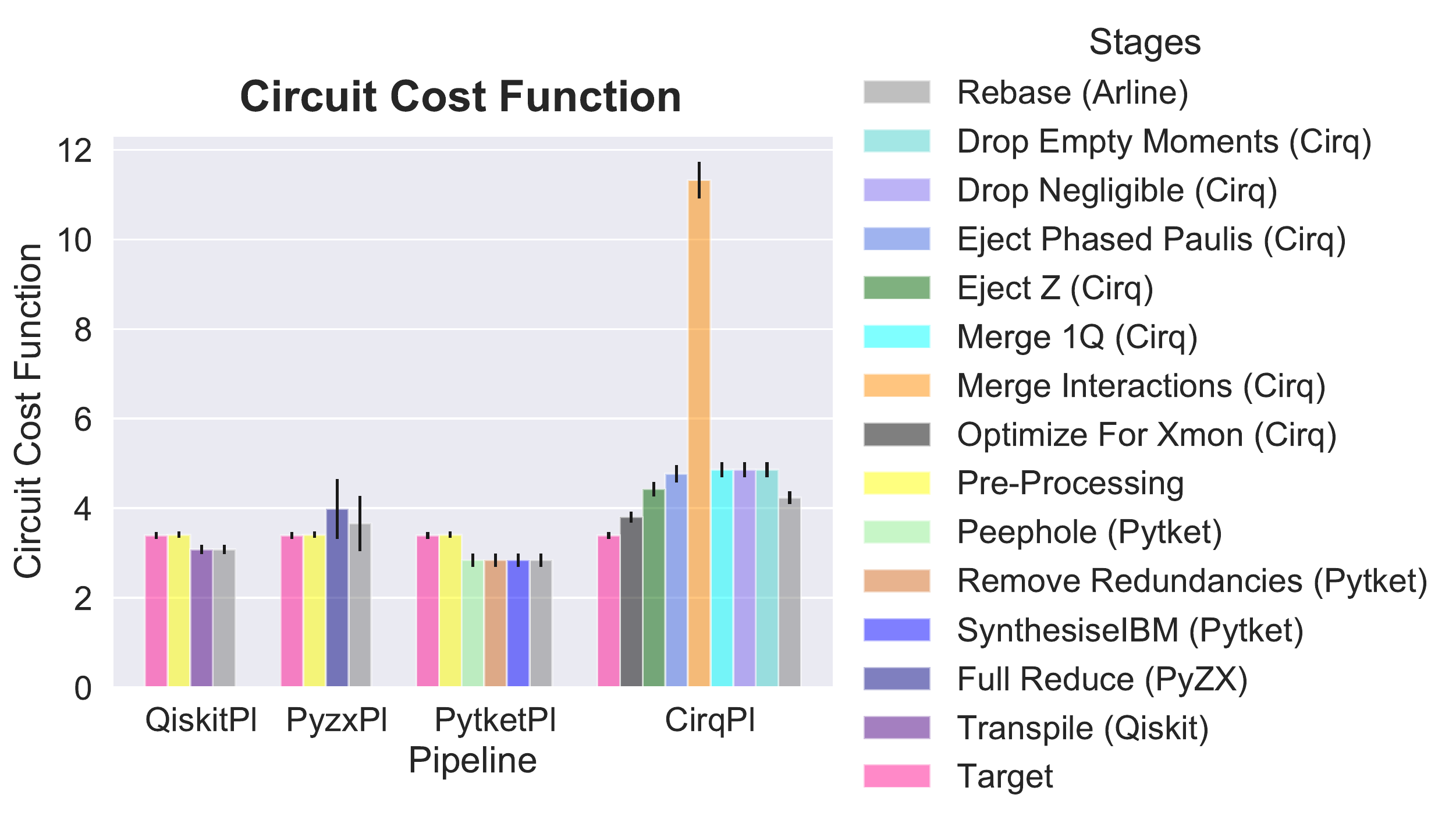}}
        \caption{Evolution of circuit metrics across compilation pipeline. (Left panel) Single-qubit gate count, (Middle panel) two-qubit gate count (Right panel) circuit cost function for each compilation stage, the stages within each pipeline corresponds to default pipeline configuration, Table \ref{table:pipelines}. \textit{MergeInteractions (Cirq)} subroutine results in a significant increase of single-qubit gates at the corresponding compilation stage, that boosts the value of the circuit cost function. The target circuits correspond to (ii): $[CX,U_3]$ random circuits  with high $CX$ density. Hardware parameters for the circuit cost function correspond to Mock IBM All2All 10q device, see Table \ref{table:cost_func_params}.}
        \label{fig:random_cx_u3}
    \end{figure}

    Arline Benchmark platform allows to analyse the transformation of circuit metrics across compilation pipeline between individual stages, see Fig. \ref{fig:random_cx_u3}. This allows to diagnose problems or inefficiencies due to a specific compilation subroutine. For example, from Fig. \ref{fig:random_cx_u3} we can see that \textit{MergeInteractions (Cirq)} subroutine increases $gc(1Q)$ by an order of magnitude, which signals about the non-optimal algorithmic implementation of this subroutine.
    
\begin{figure}[H]
        \resizebox{\textwidth}{!}{
        \includegraphics{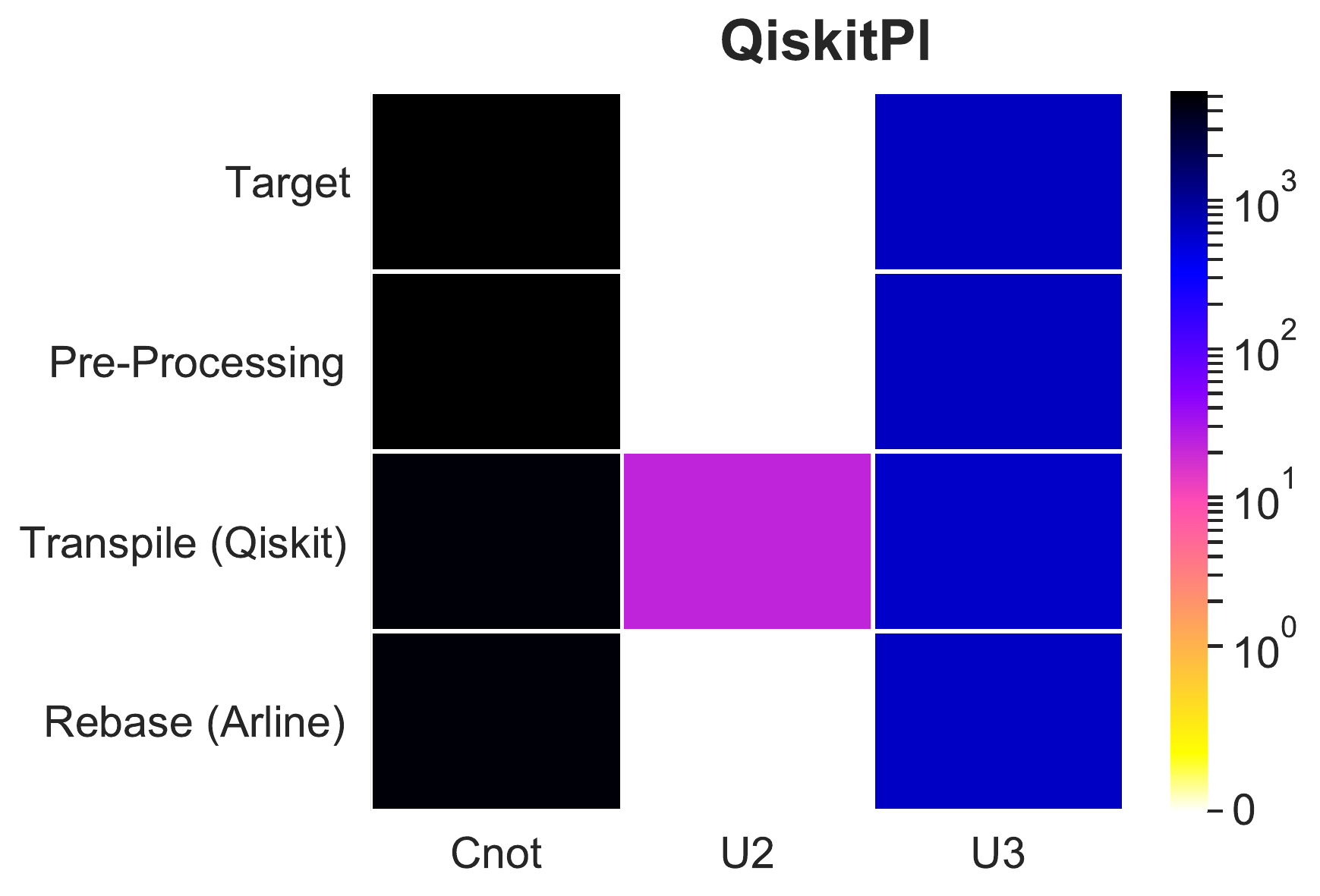}
        \includegraphics{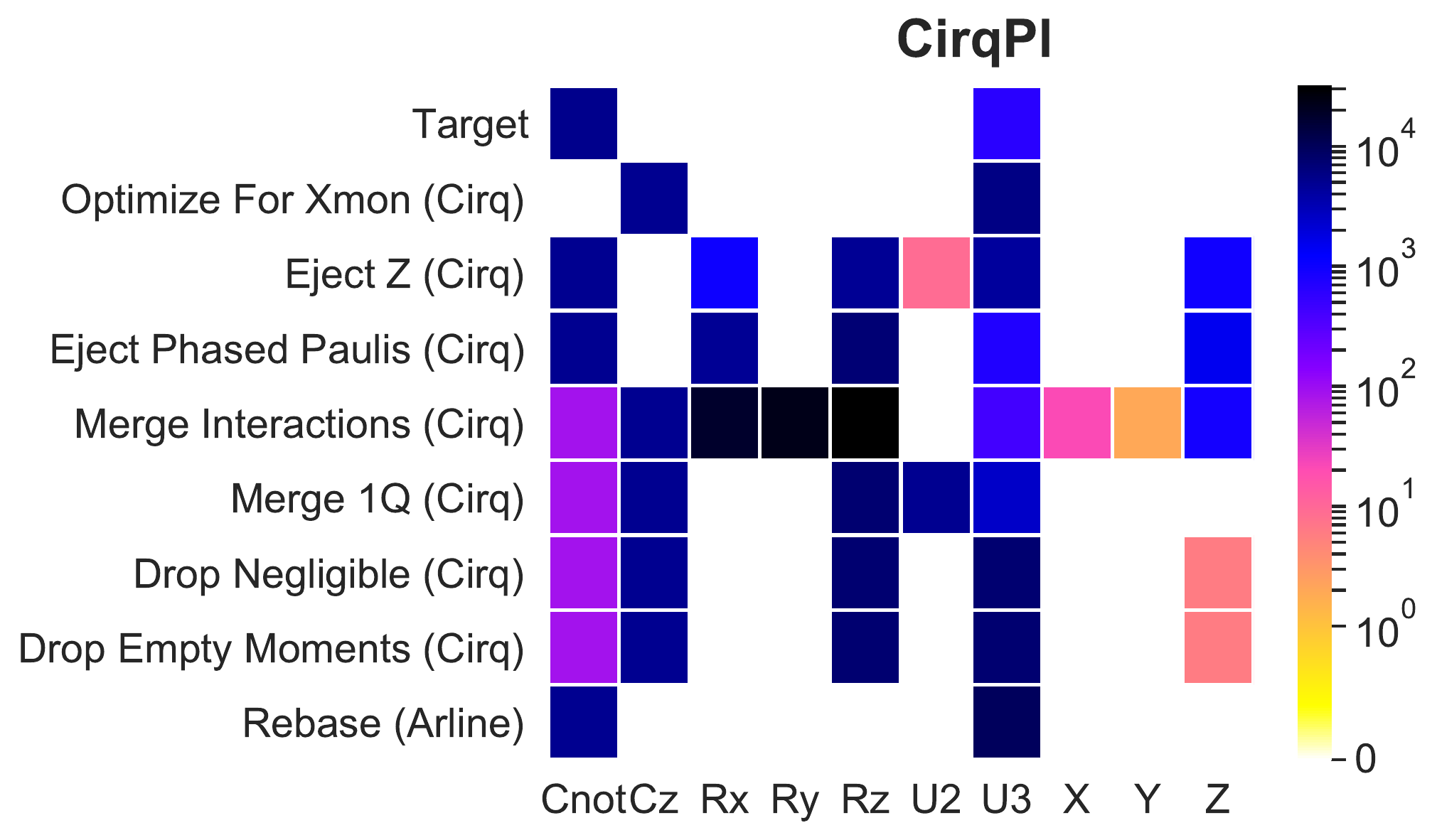}
        \includegraphics{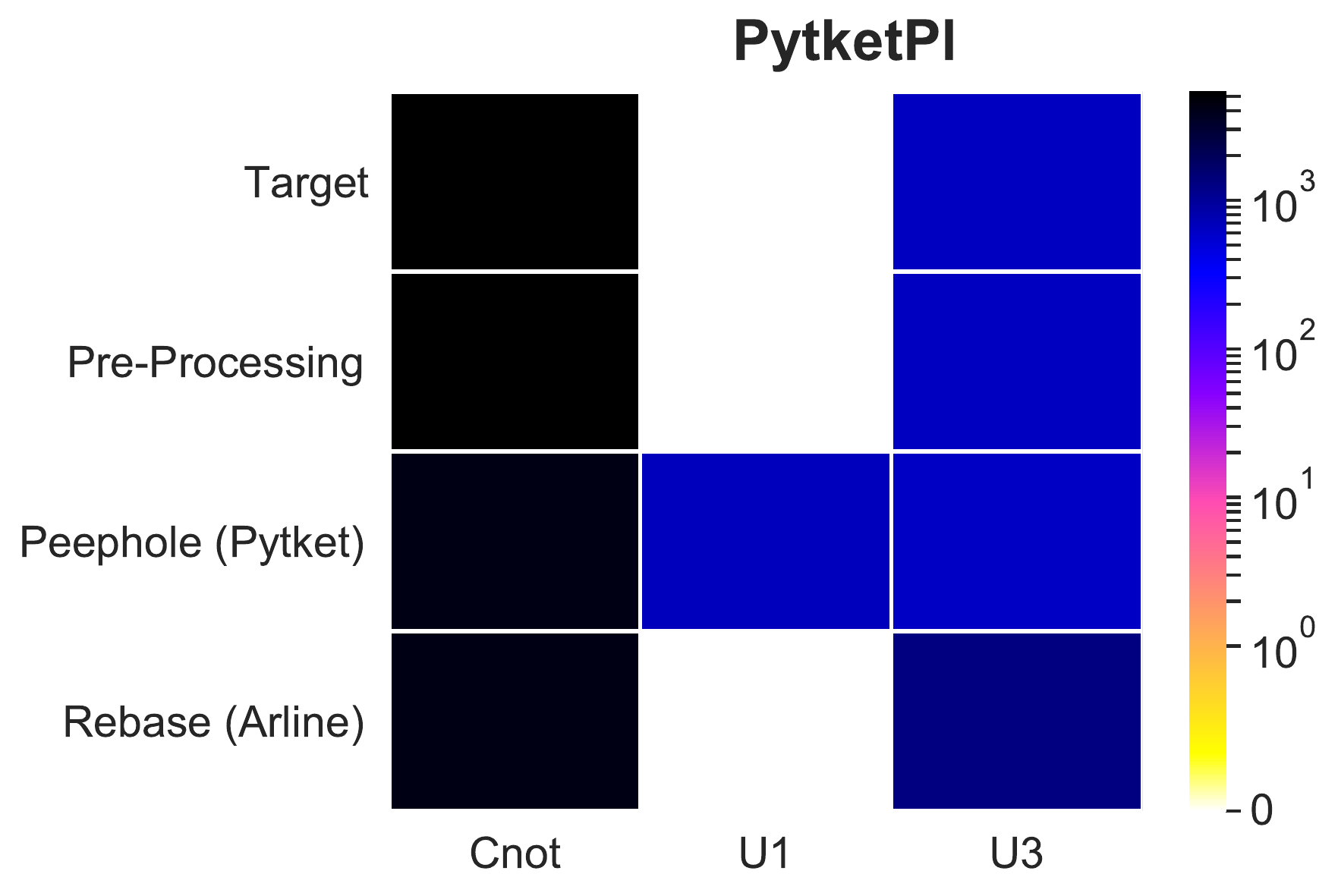}
        \includegraphics{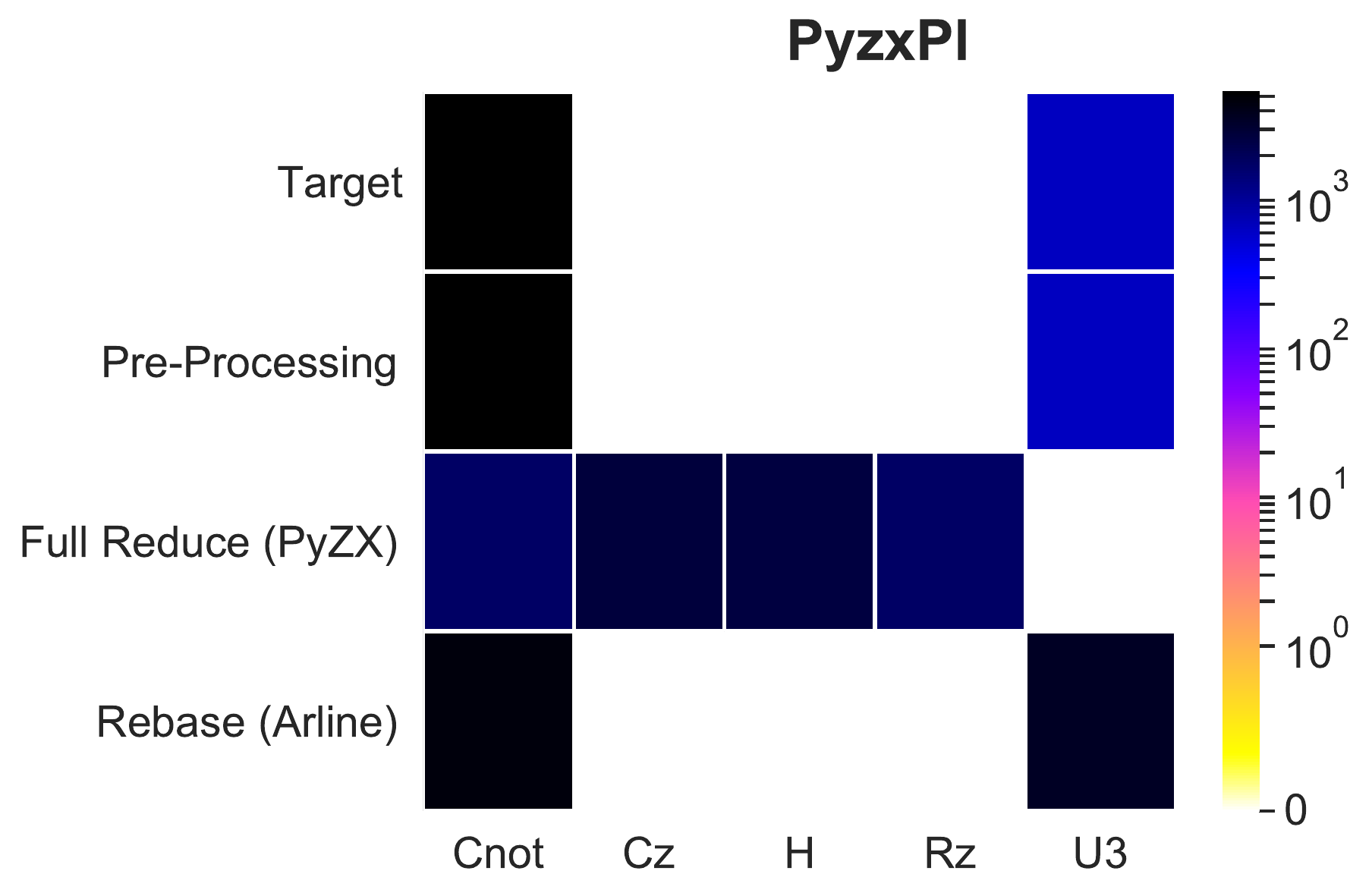}}
        \caption{Visualization of the gate set transformation across different stages of compilation pipeline. Color shows gate count for each gate type in four compilation piplines:  QiskitPl,  CirqPl,  PytketPl and  PyZXPl. Target circuits correspond to random $[CX, U_3]$  circuits of type (ii). 
        The final compilation stage \textit{Rebase (Arline)} corresponds to rebase to $[CX, U_3]$ output gate set.}
        \label{fig:gate_composition}
    \end{figure}

    In Fig. \ref{fig:gate_composition} we visualize transformations of gate sets between stages of the compilation pipeline. In order to perform circuit optimizations subroutines perform conversion to a favourable gate set.
    An auxiliary  \textit{Pre-Processing (Arline)} stage adds a barrier gate followed by terminal measurement gates. The addition of measurement gates is important to keep track of logical qubit permutations introduced by the compiler at the end of the circuit.
            
\begin{figure}[H]
        \resizebox{\textwidth}{!}{\includegraphics{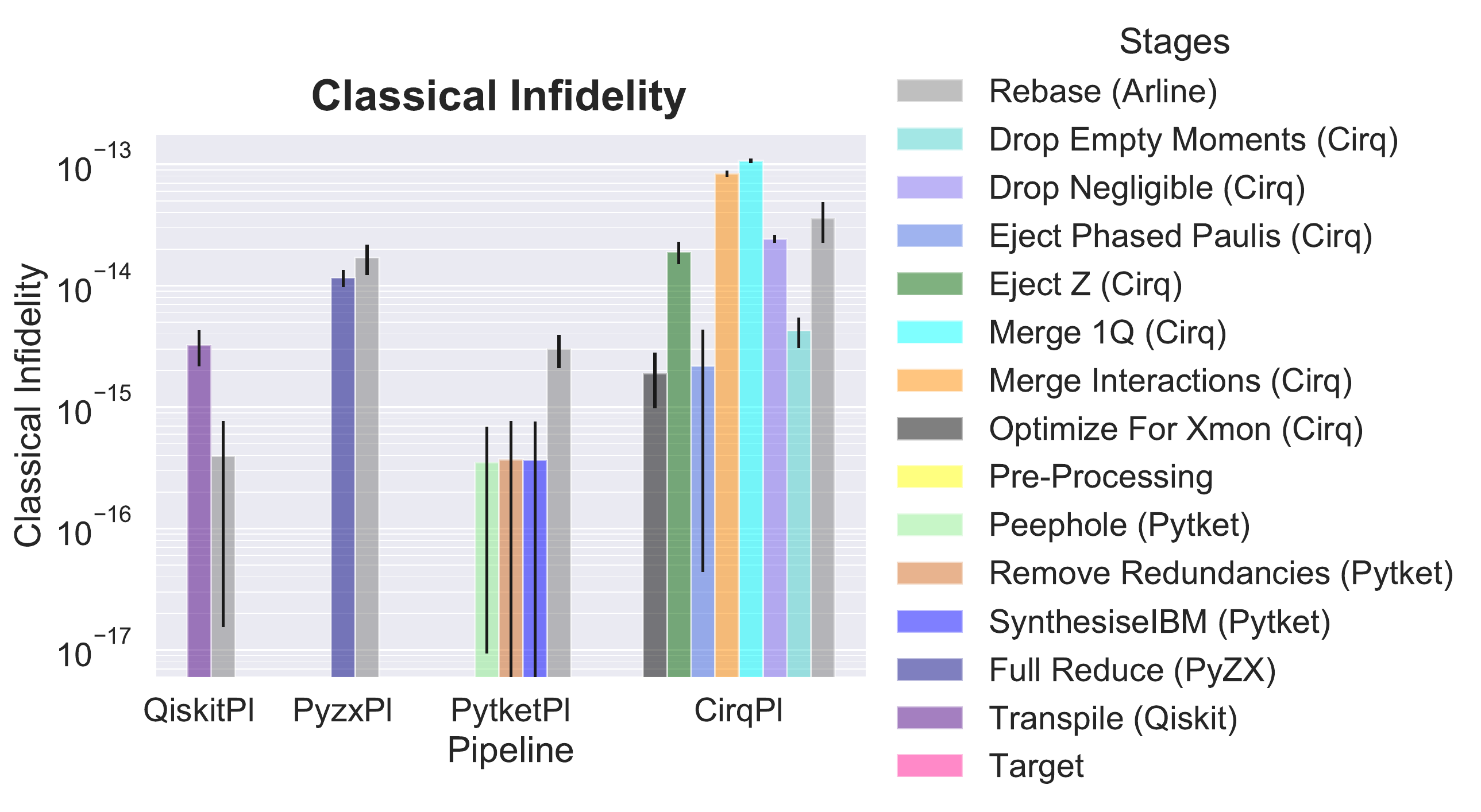}
        \includegraphics{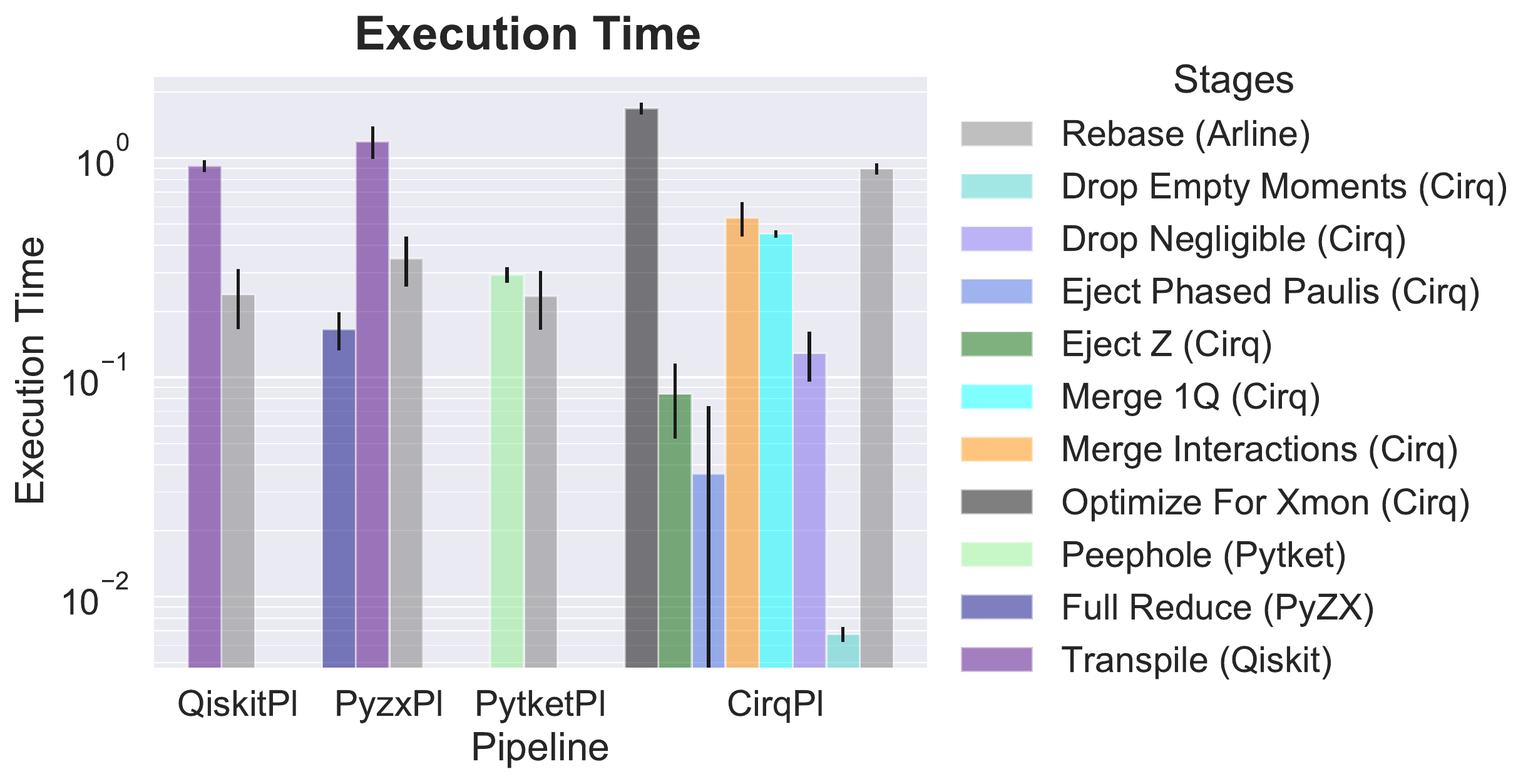}
        \includegraphics{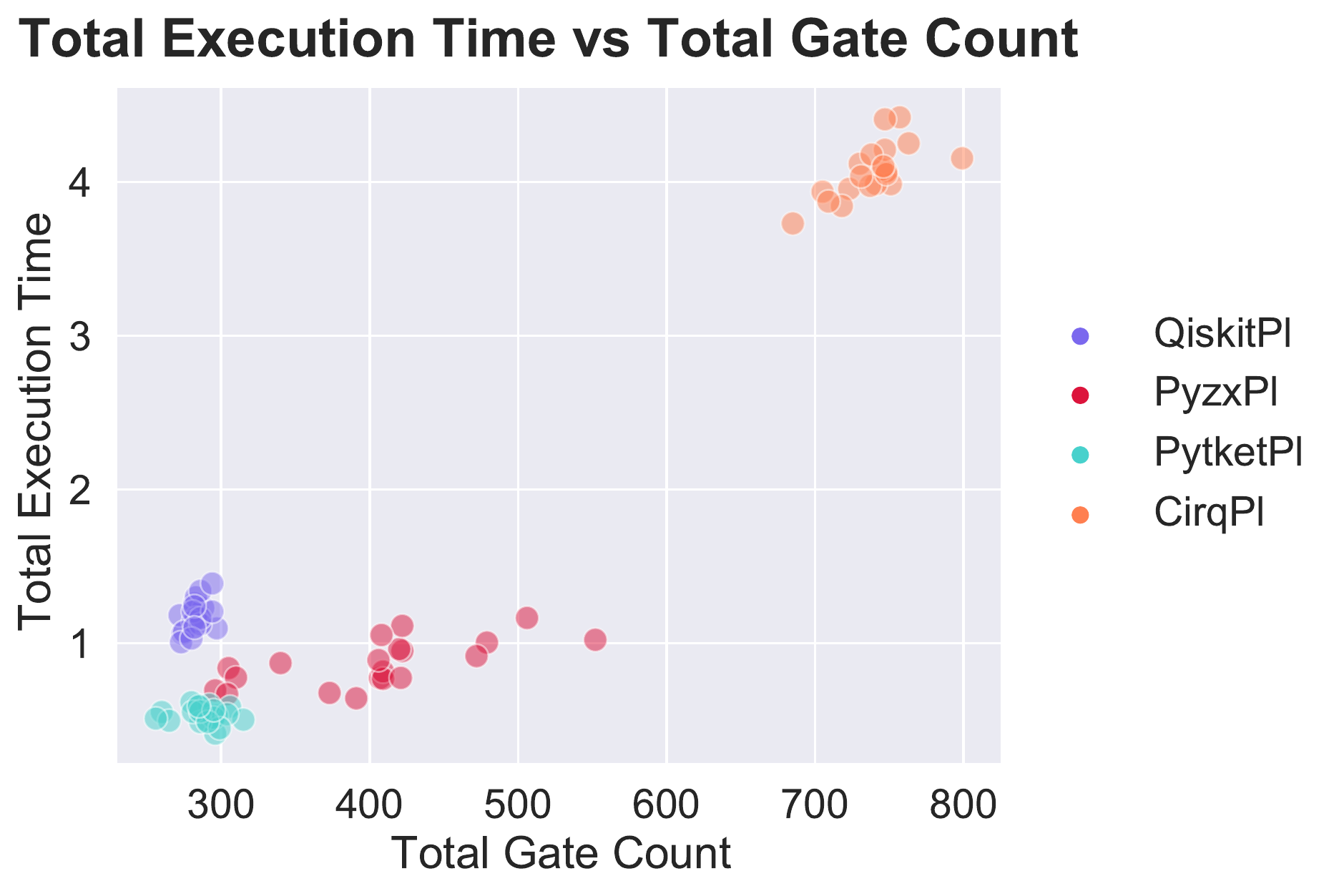}}
        \caption{(Left panel) Circuit equivalence checking by direct computation of classical fidelity $|1- \mathcal F_{cl}|$ between the target and compressed circuits, see Eq. \ref{eq:F_cl}. The deviation of $\mathcal F_{cl}$ from unity is on the level of machine precision for all compilation frameworks. (Middle panel) Runtime (seconds) for each compilation stage spent by the subroutine. (Right panel) Total compilation runtime vs  total gate count of the output circuit. Target circuits correspond to random $[CX,U_3]$ circuits (ii).}
        \label{fig:infidelity}
    \end{figure}

    Execution time is another important characteristics of a quantum compiler, especially having in mind that compilation of large quantum circuits can be prohibitively slow. Arline Benchmarks collects information about time spent by each compilation stage, see Fig. \ref{fig:infidelity}(b). We found that Pytket with the backend implemented in C++ showed the best runtime performance. Qiskit, PyZX and Cirq fully implemented in Python have comparable execution times. Low latency compilation becomes critical for variational based quantum algorithms such as QAOA and VQE, where the variational single-qubit gate angles are updated online. This problem can be alleviated by performing parametric (symbolic) compilation of variational quantum algorithms. Qiskit and Pytket frameworks support compilation of parametric quantum circuits, and parametric circuits are supported in Cirq's Tensorflow Quantum extension. In the present paper, we consider only non-parametric compilation, although benchmarking of parametric circuit compilation could be introduced in later versions of Arline Benchmarks platform.
    
    \subsection{Structured Circuits: Quantum Algorithms} \label{sec:structured_circ_compression}

    In this section, we analyse the performance of quantum compiler frameworks on structured input circuits, corresponding to four classes of quantum algorithms (see Table \ref{tab:quantum_algos}): 
    \begin{itemize}
        \item[(1)] Unitary coupled cluster  circuits (UCCSD) for electronic structure calculations with variational eigensolver algorithm (VQE);
        \item[(2)] Trotterized quantum dynamics of transverse field Ising model with local and long-range interactions;
        \item[(3)] Quantum algorithm for European option payoff calculation based on the Amplitude Estimation Algorithm;
        \item[(4)] Grover search in an unstructured database.
    \end{itemize}
    We do not consider arithmetic reversible circuits (adders, ALUs, integer modulo functions, etc.) commonly used for benchmarking of quantum circuit optimizers, see e.g. Refs.  \cite{nam2018automated, quilc2020, hietala2021verified}, which are more appropriate to study in the context of fault-tolerant quantum computing rather than for NISQ applications. Similar to Sec. \ref{sec:random_circ_compression} here we focus on hardware with all-to-all connectivity for the purpose of separating the effects of circuit optimization and overhead introduced by routing/mapping. Optimization of structured circuits on hardware with restricted connectivity is considered in Sec. \ref{sec:dedicated_hardware}. 
    
\begin{figure}[H]
        \centering
        \includegraphics[scale=.65]{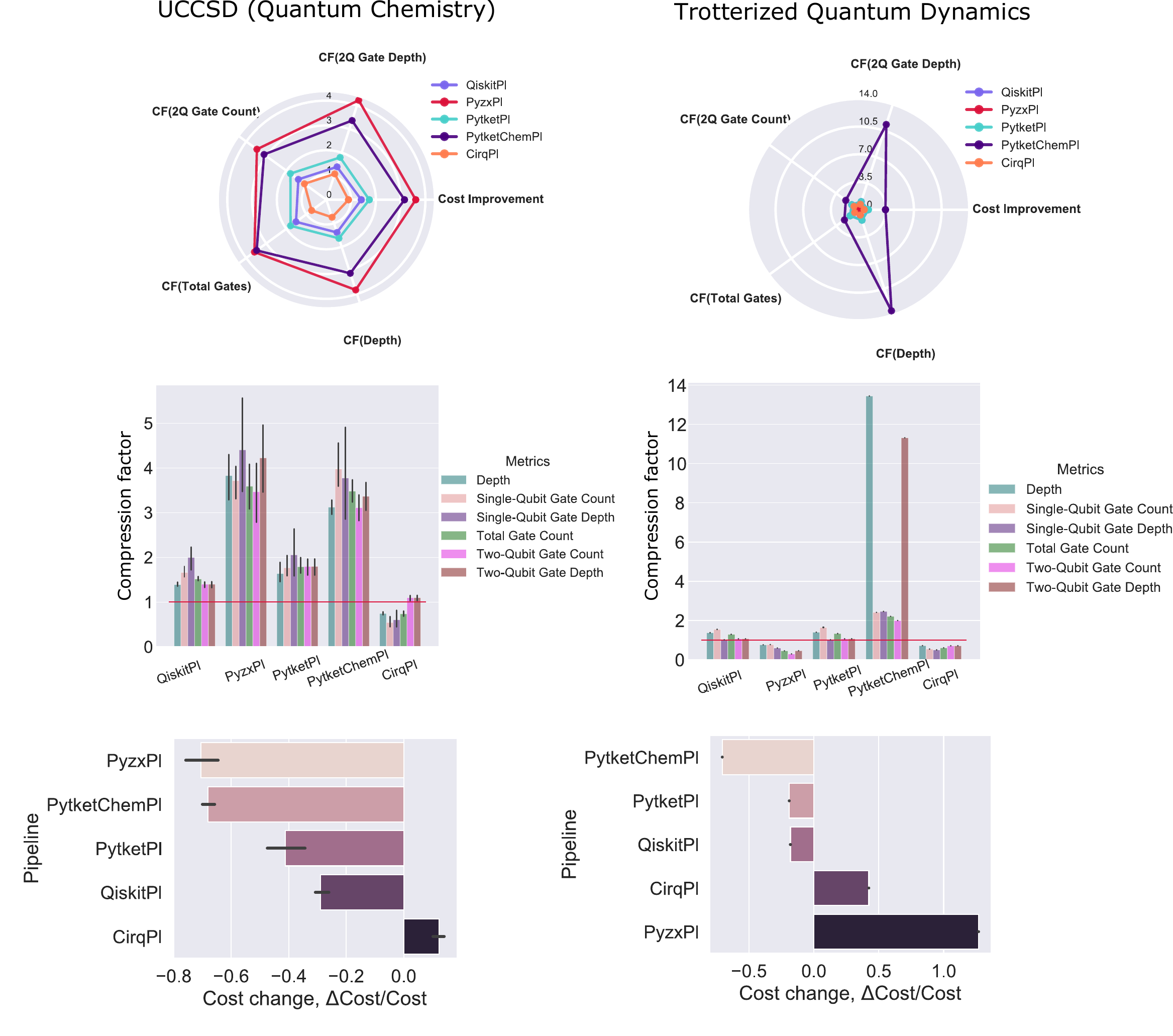}
        \caption{Optimization of structured quantum circuits on fully connected hardware. Four quantum algorithms are considered, see Table \ref{tab:quantum_algos}: UCCSD circuits for VQE algorithm (H$_2$, H$_2$O, LiH, NH molecules), Trotterized quantum dynamics for an Ising spin chain in the transversal magnetic field. The compression factor is evaluated between a preprocessed circuit, where all gates are converted to $[CX, R_z, R_x]$ gate set and a final circuit rebased to $[CX, U_3]$ gate set. Input circuit QASM files could be found in \cite{ArlineBenchmarksRepo}. Bottom row: circuit cost function change after optimization, $(\mathcal{C}_{out}-\mathcal{C}_{in})/\mathcal{C}_{in}$.}
        \label{fig:compression_struct_all2all}
    \end{figure}

    The summary of the benchmarking results for four types of quantum algorithms are presented in Fig. \ref{fig:compression_struct_all2all}. Along with compilation pipelines considered in the previous sections (\textit{QiskitPl}, \textit{PytketPl}, \textit{CirqPl}, \textit{PyZXPl}), we add a dedicated pipeline \textit{ChemPassPl (Pytket)} introduced in \cite{tket2020} specifically for optimization of VQE-like circuits. The pipeline \textit{ChemPassPl} contains \textit{PauliSimp (Pytket)} subroutine that performs non-local resynthesis of Pauli exponential blocks (phase gadgets), mimicking the functionality of ZX-calculus-based circuit optimization. First, from Fig. \ref{fig:compression_struct_all2all} we see that \textit{QiskitPl} and \textit{PytketPl} demonstrate similar compression performance on average successfully improving the circuit cost function.
    
    \begin{figure}[H]
        \centering
        \resizebox{\textwidth}{!}{\includegraphics{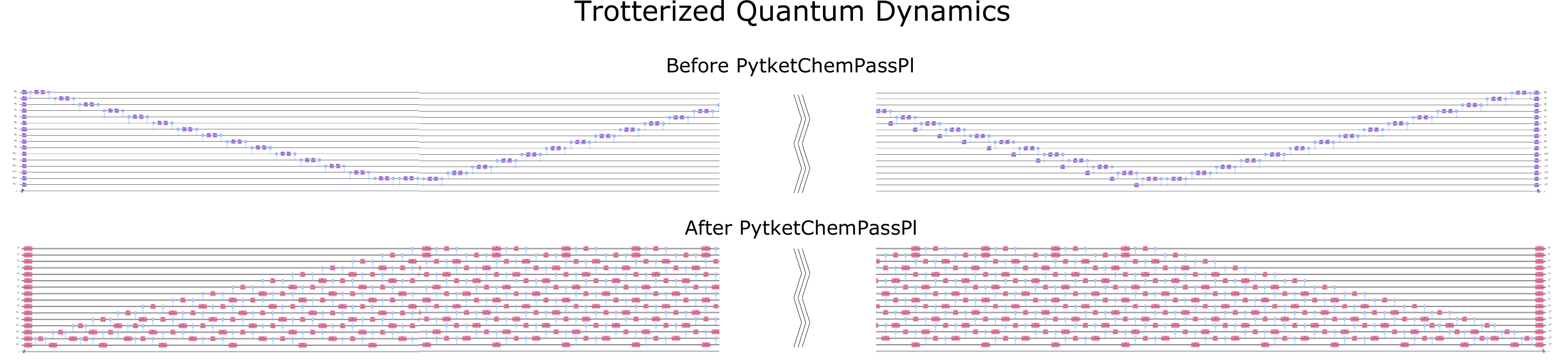}}
        \caption{Fragment of a trotterized circuit for quantum dynamics benchmark (transverse Ising spin chain) before/after optimization with \textit{PytketChemPl} pipeline. (Top panel) The input circuit is sparse and contains 20 identical Trotter steps. (Bottom panel) Optimized circuit by \textit{PytketChemPassPl}
        has a dense layout with significantly reduced depth and $CX$ depth.}
        \label{fig:circuit_pattern}
    \end{figure}
    
    Compression metrics for each quantum algorithm class were averaged over several circuit instances as shown in Fig. \ref{fig:compression_struct_all2all}. In our experiments, we found that compression performance of quantum compilers remain consistent within each circuit class. The compression ratios $CF$ for the metrics of interest have a low variance around the mean values, see Fig. \ref{fig:compression_struct_all2all}, despite of large variations in the size of the input circuits (e.g. total gate counts of UCCSD circuits used for benchmarking differ by two orders of magnitude). 
    
    For the case of the circuits implementing Trotter product formula for quantum dynamics simulation, we found that \textit{ChemPassPl} significantly outperformed other frameworks in terms of depth and $CX$ depth reduction, see the right column in Fig. \ref{fig:compression_struct_all2all}. Interestingly, although the depth-dependent compression ratios for \textit{ChemPassPl} is an order of magnitude larger compared to competitors, $CF(\textrm{depth})\sim 10$, the reduction in $CX$ gate count is not as dramatic, $CF(gc,CX)\sim 2$,  demonstrating that \textit{ChemPassPl} modified the gate pattern in the target circuit by drastically increasing gate densities in the output circuit, see Fig. \ref{fig:circuit_pattern}. For other circuit classes, the change of the circuit layout after circuit optimization is harder to interpret. 
    
    Interestingly,  \textit{PyZXPl} and \textit{ChemPassPl}   demonstrated a similar compression performance for UCCSD circuits,
    see Fig. \ref{fig:compression_struct_all2all} (first column).
    This is not entirely surprising provided that both circuit compression strategies
    rely on ZX-calculus-based rewriting rules. 
    However, for the Trotterized quantum dynamics benchmark  \textit{PyZXPl} significantly increases output circuit size as shown in the second column in Fig. \ref{fig:compression_struct_all2all}. Despite the fact that both UCCSD and Trotter circuits have a recurrent structure, the performance of the PyZX optimizer on these circuit classes is drastically different.

    It is natural to ask a question whether one can predict the potential for further compression for a given circuit instance. In a general case, this problem is likely to be QMA-hard.
    In our experiments, we did not find a simple method to predict the compression performance of a given compilation framework on a specific class of quantum algorithms. In order to forecast compression potential for a given quantum circuit, one would need to know  specific gate patterns each  subroutines is sensitive to and search for such patterns in the given circuit. Although for some simple optimization subroutines this analysis is quite straightforward (e.g. KAK-based compression of two-qubit subcircuits in a larger circuit), for other subroutines there are no existing tools to perform such analysis.
    A naive approach to this problem could be to analyse properties of the commutativity graph of the quantum circuit (so-called canonical DAG representation of quantum circuits \cite{iten2019exact}). However,  the information about the local commutativity structure of the circuit could be insufficient for predicting compression performance of optimization strategies which are very non-local (e.g. \textit{PauliSimp} and \textit{FullReduce} optimization subroutines based on ZX-calculus).
    
    

    \begin{figure}[H]
        \centering
        \includegraphics[scale=.65]{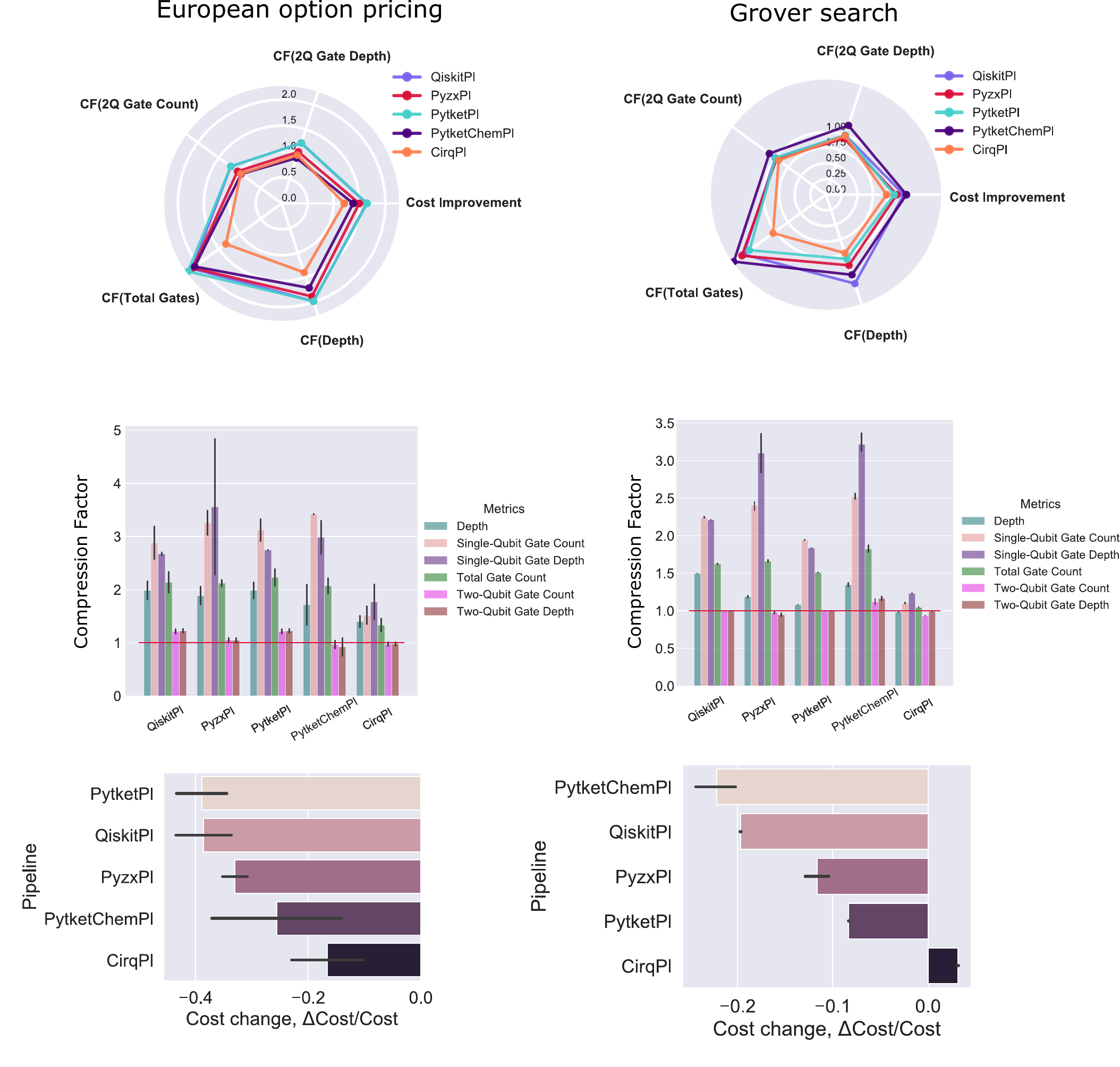}
        \caption{Optimization of structured quantum circuits on fully connected hardware. Here we consider two quantum algorithms: option pricing algorithm via amplitude estimation (European put/call option), Grover search algorithm, see Table \ref{tab:quantum_algos}.}
    \end{figure}
    
    \subsection{Structured vs Randomized Circuits}
    
    In the previous sections, we considered compression performance of quantum compilers for random input circuits and input circuits corresponding to quantum algorithms. It is interesting to ask a question whether random or structured quantum circuits have more potential for circuit optimization. The results of experiments in  Sec. \ref{sec:random_circ_compression} and Sec. \ref{sec:structured_circ_compression} can not be directly compared since the gate sets and the densities of single-qubit and two-qubit gates were different in both test cases. In order to address the question, we compared the compression performance of quantum compilers for the original structured circuits and structured circuits with randomized gate ordering/gate placement. When performing randomization of structured circuits we keep the same gate composition (the total number of gates of each type) as well as the angles parametrizing single-qubit gates unchanged. We consider two types of circuit randomization procedures:  
    \begin{itemize}
        \item[(i)] Circuits with fully randomized order and placement of gates (``fully randomized'' circuits). The indexes of $n$-qubit gates were sampled using the procedure similar to the one  described in Sec. \ref{sec:random_circ_gen}: the indexes ($i_1, \ldots, i_n$) of a gate $g_{i_1, \ldots, i_n}$ are chosen at random from all possible combinations with equal probabilities (the number of possible placements of $n$-qubit gate on $N$ qubits is $C_N^n$).
        \item[(ii)] Circuits with shuffled order of gates, such that the indexes of qubits the gate acts on remain unchanged $g_{i_1, \ldots, i_n}$, but the order of gates in the circuit is randomized (denoted as ``shuffled'' circuits). 
    \end{itemize}
    
    The procedure for generating fully randomized circuits (i) is similar to the method used for generation random circuits in Sec. \ref{sec:random_circ_compression}, while the shuffled circuits of type (ii) partially keep the layout of original circuits. The results of the experiment are presented in Table \ref{tab:shuffled}. We found that full randomization of structured circuits decreases compression factors $CF(gc, 1Q)$ and $CF(gc, 2Q)$ and therefore negatively impacts circuit compression performance. This effect is especially pronounced when considering domain specific compilation subroutines/pipelines which are sensitive to distinct patterns in the input circuits, e.g. \textit{PyZXPl} and \textit{PytketChemPl} for UCCSD circuits demonstrate significant decline in compression performance after full randomization or gate reshuffling. With the exception of  domain-specific \textit{PyZXPl} and \textit{PytketChemPl} compilation pipelines, gate re-shuffling increases $CF(gc, 2Q)$ for remaining pipelines \textit{QiskitPl}, \textit{PytketPl} and \textit{CirqPl} for each class of quantum algorithms considered. This could be put in contrast to circuits, obtained via full randomization where compression metrics declined compared to compression metrics for original structured circuits. Such behaviour can be explained due to gate shuffling procedure creates trivial identity subcircuits with nonzero probability, e.g. $(CX_{ij})^2 =1$, $H_i^2 = 1$, etc., which are easily identified and removed by circuit optimization subroutines in \textit{QiskitPl}, \textit{PytketPl} and \textit{CirqPl}.
    
    \begin{table}[H]
        \caption{Compression ratios (higher is better) for circuits corresponding to four classes of quantum algorithms (original circuits), as well as compression ratios for randomized circuits generated from the input circuits using two types of randomization procedures (``fully randomized'' and ``shuffled''). All-to-all hardware connectivity is assumed. Highlighted entries correspond to the best compression within each circuit class according to $CX$ gate count.}
        \centering
        \resizebox{\textwidth}{!}{\begin{tabular}{|l|c|c|c|c|c|}
            \hline
            \multicolumn{1}{|c|}{\textbf{Circuit Type}} & \textbf{Pipeline} & \textbf{VQE (UCCSD)} & \textbf{Finance (Option Pricing)} & \textbf{Grover Search}& \textbf{Trotter} \\
            \hline
            \multirow{5}{*}{$CF(gc, CX)$, Original} & QiskitPl & 1.40 & \textbf{1.21} & 1.00 & 1.07\\
                        & PytketPl & 1.79 & \textbf{1.21} & 1.00 & 1.07\\
                        & PytketChemPl & 3.11 & 0.965 & \textbf{1.19} & \textbf{2.00}\\
                        & PyZXPl & \textbf{3.46} & 1.05 & 0.98 & 0.30\\
                        & CirqPl & 1.03 & 0.97 & 0.94 & 0.71 \\
            \hline
            \multirow{5}{*}{$CF(gc, CX)$, Fully Randomized} & QiskitPl & 1.12  & 1.02  & 1.08  & 1.03 \\
                        & PytketPl &  1.32 & \textbf{1.05}  &  \textbf{1.16} &  \textbf{1.05}\\
                        & PytketChemPl & 1.20 & 0.17  & 0.28  &  0.13 \\
                        & PyZXPl & \textbf{1.82}  & 0.45  & 0.61  & 0.54 \\
                        & CirqPl & 1.08  & 0.94  & 1.05  & 1.00 \\
            \hline
            \multirow{5}{*}{$CF(gc, CX)$, Shuffled} & QiskitPl & 1.75 & \textbf{1.46} & 1.36 & \textbf{1.77}\\
                        & PytketPl & 1.98 & 1.43 & \textbf{1.58} & 1.75\\
                        & PytketChemPl & 1.35  & 0.67 & 0.32  & 0.27  \\
                        & PyZXPl & \textbf{2.53} & 0.91 & 0.90 & 1.35\\
                        & CirqPl & 1.62 & 1.01 & 1.24 & 1.49\\
            \hline
        \end{tabular}}
        \label{tab:shuffled}
    \end{table}

    \subsection{Circuit optimization for a dedicated hardware}
    \label{sec:dedicated_hardware}
    
    In this section, we consider an entire circuit compilation pipeline consisting of circuit optimization, qubit mapping and routing stages for hardware architectures with restricted connectivity. In this test we combine benchmarking analysis of Sections \ref{sec:routing} and \ref{sec:compression}, the list of considered compilation pipelines is shown in the last column of Table \ref{table:pipelines}. Since the implementation of PyZX module does not support qubit routing, we incorporated SABRE routing method from \textit{Transpile (Qiskit)} as a part of \textit{PyZXPl} pipeline. 
    
\begin{figure}[H]
        \centering
        \resizebox{\textwidth}{!}{\includegraphics{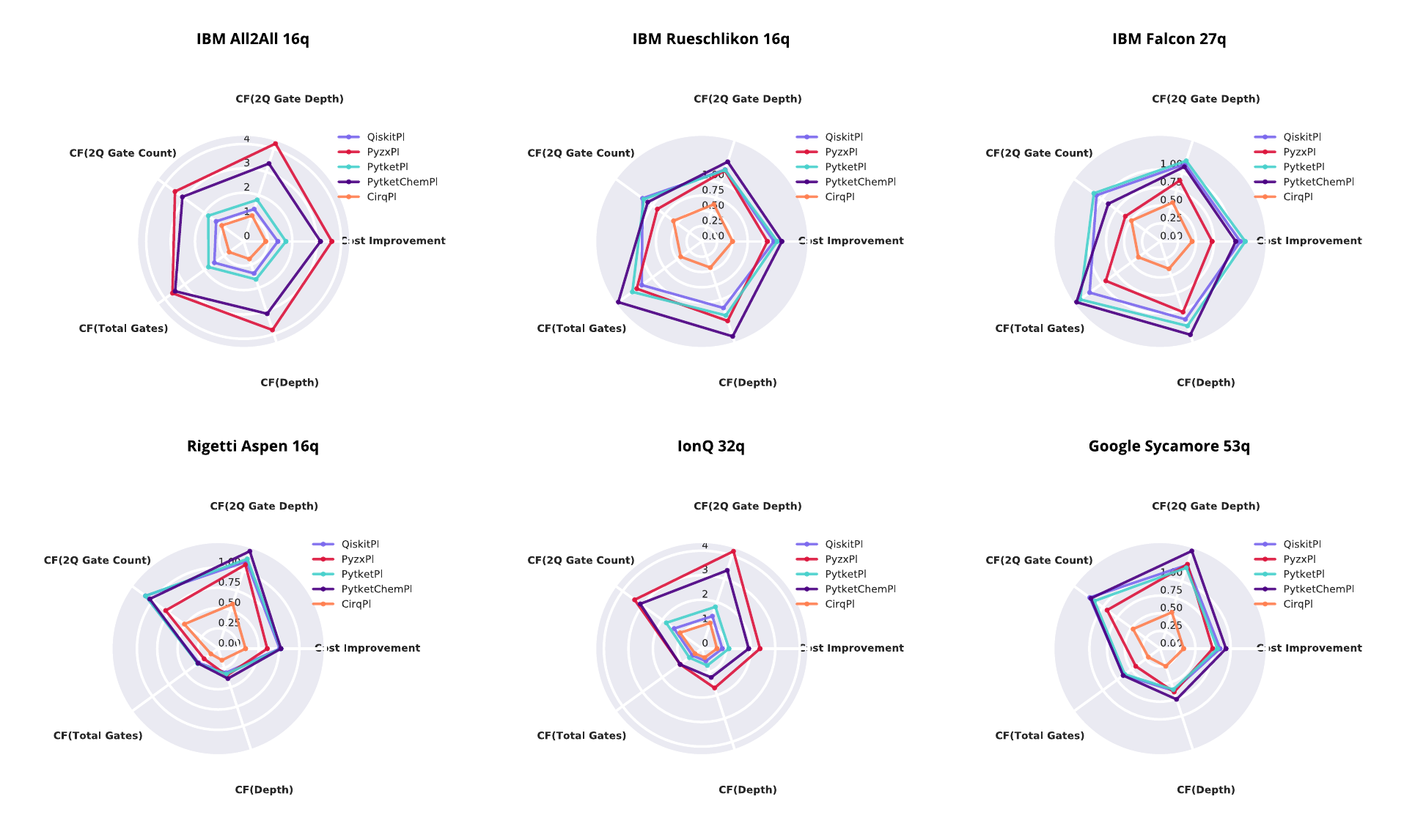}}
        \caption{Compression ratios for VQE UCCSD circuits on six hardware architectures (higher is better). Top row: IBM All2All 16q,  IBM Rueschlikon 16q,  IBM Falcon 27q; Bottom row: Rigetti Aspen 16q,  IonQ 32q,  Google Sycamore 53q. Final optimized circuits are translated to hardware native gate sets. The hardware gate sets and circuit cost parameters are listed in Table \ref{table:cost_func_params}.}
        \label{fig:structured_chem}
    \end{figure}

    In Fig. \ref{fig:structured_chem} we show circuit compilation metrics for VQE UCCSD quantum algorithm for six quantum hardware architectures: Mock IBM All2All, IBM Ruechlikon, IBM Falcon, Rigetti Aspen, IonQ, Google Sycamore (corresponding adjacency graphs are shown in Fig. \ref{fig:hardware}). We use  Mock IBM All2All device with all-to-all qubit connectivity and $[CX, U_3]$ gate set as a toy baseline hardware model. In ion trap quantum devices, such as the ones designed by IonQ the two-qubit gates are realized via interaction of ions with phonon modes, which makes it possible to physically implement all-to-all qubit connectivity~\cite{cirac1995quantum, wright2019benchmarking}. 
    
\begin{figure}[H]
        \centering
        \resizebox{\textwidth}{!}{\includegraphics{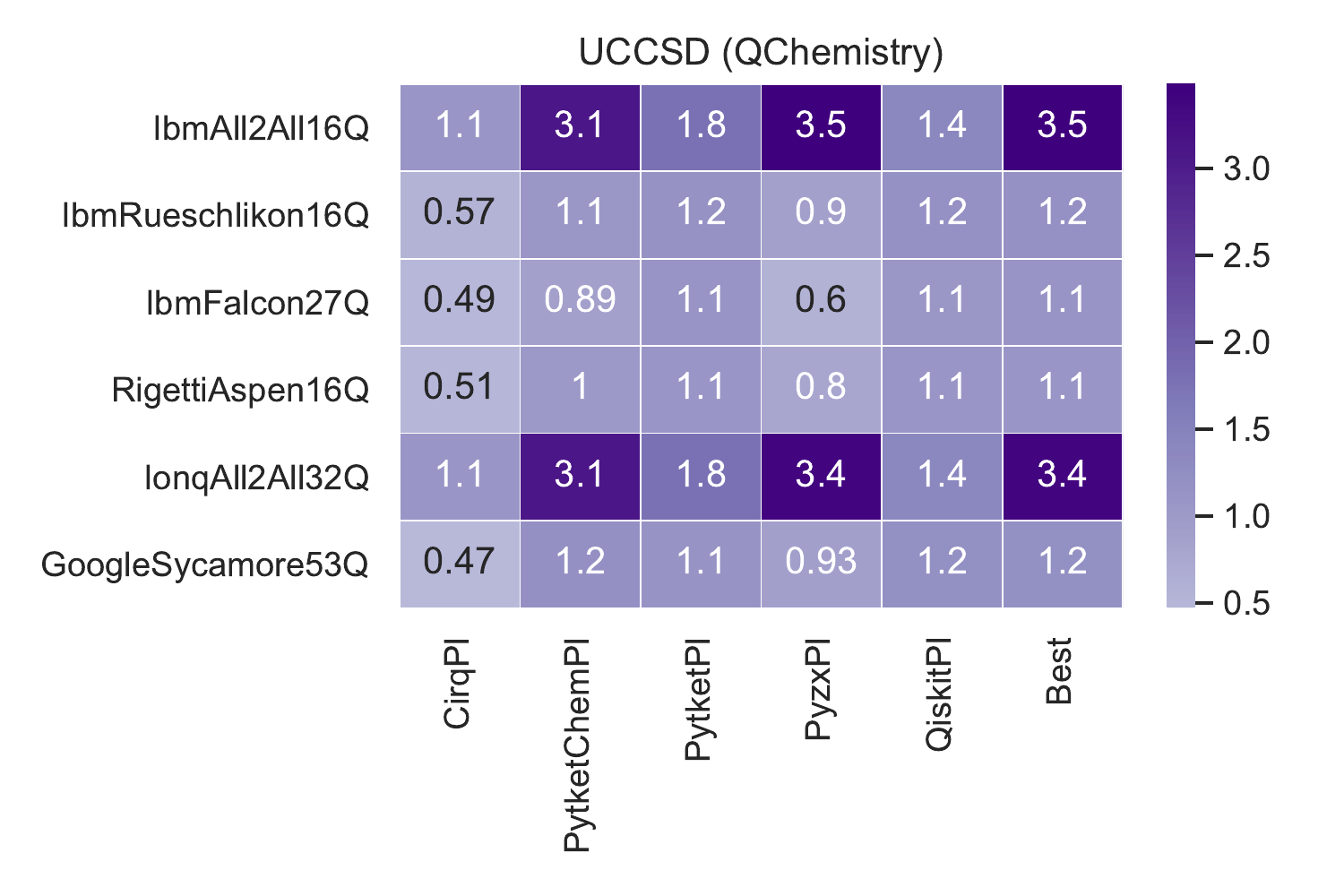}
        \includegraphics{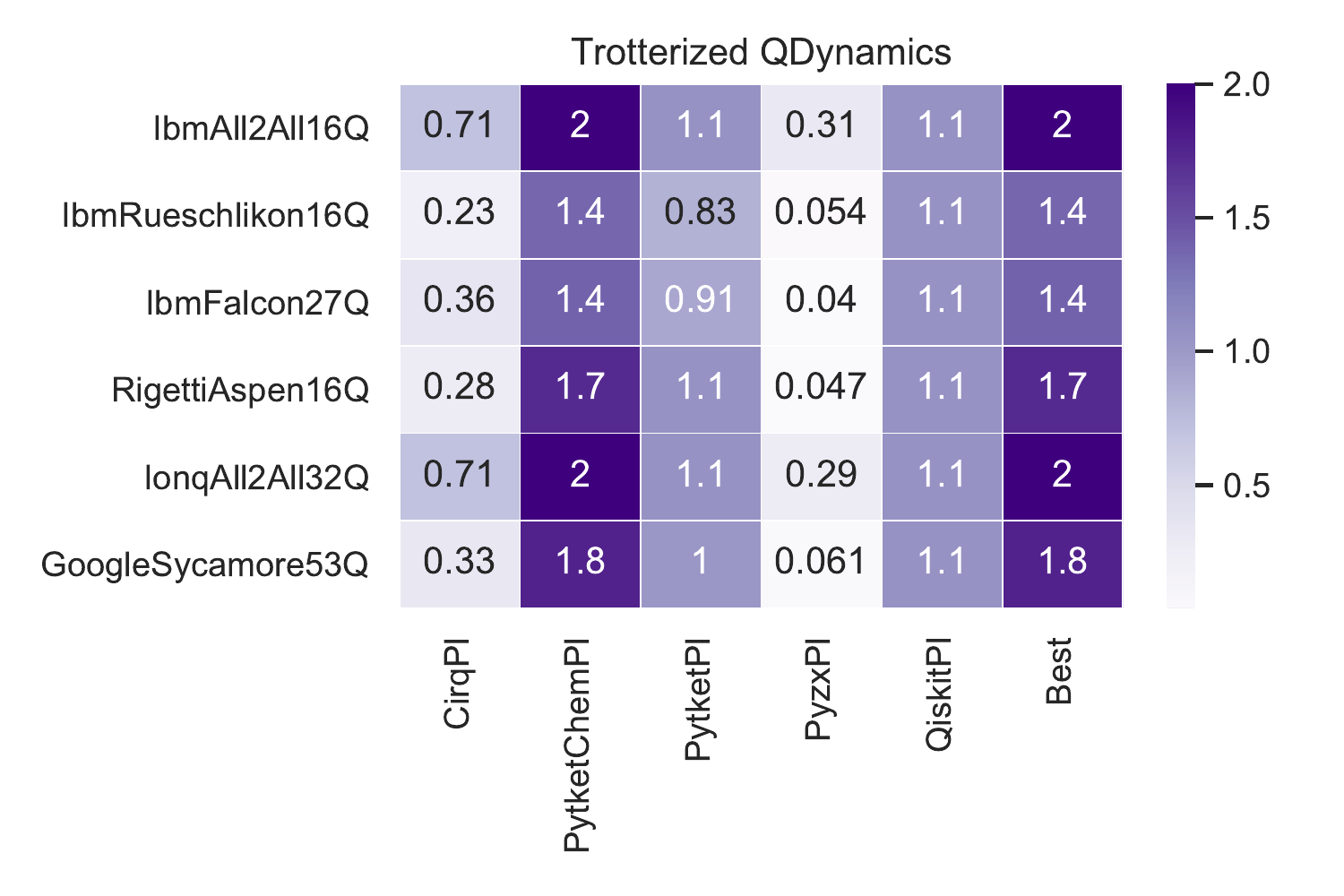}}
        \resizebox{\textwidth}{!}{\includegraphics{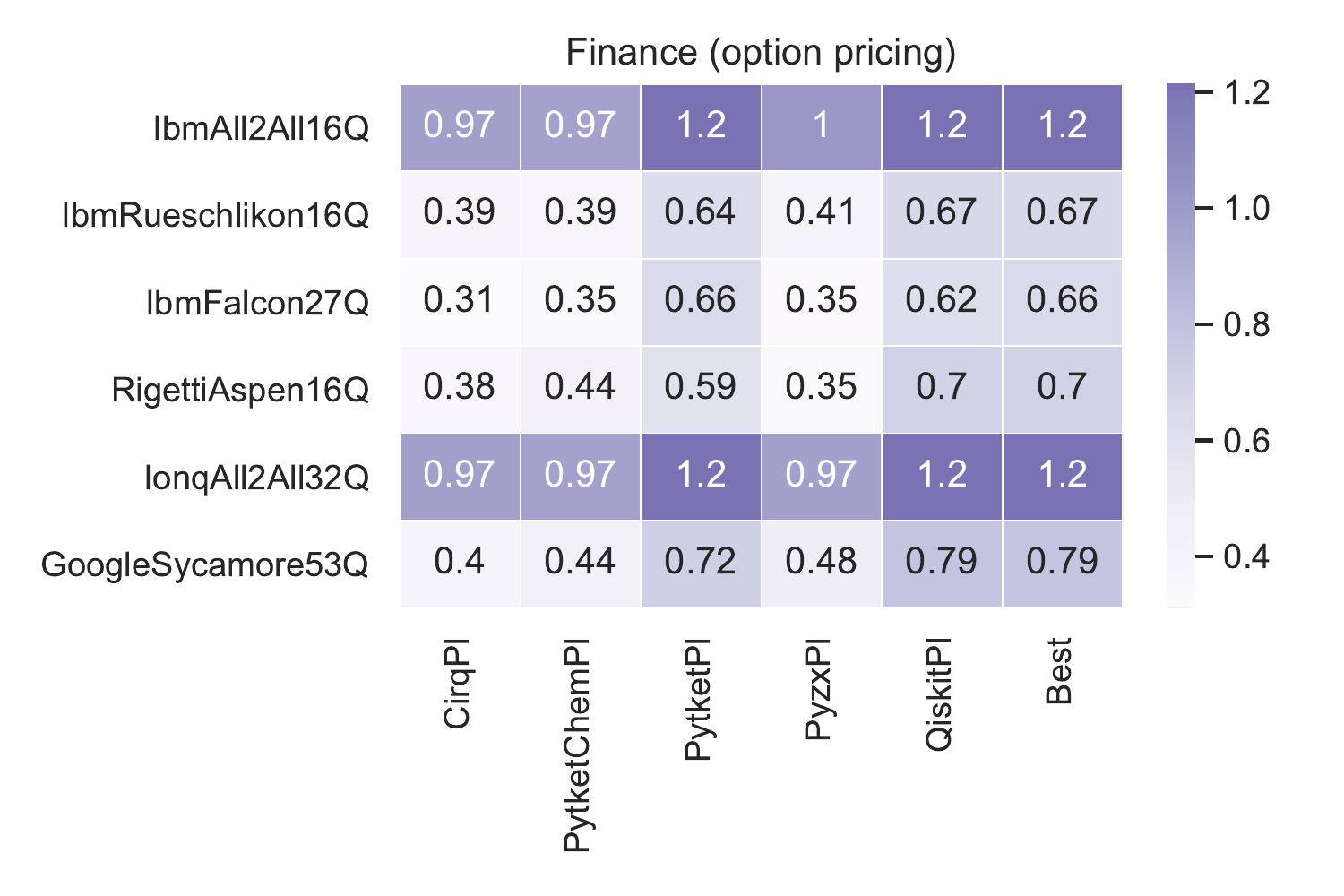}
        \includegraphics{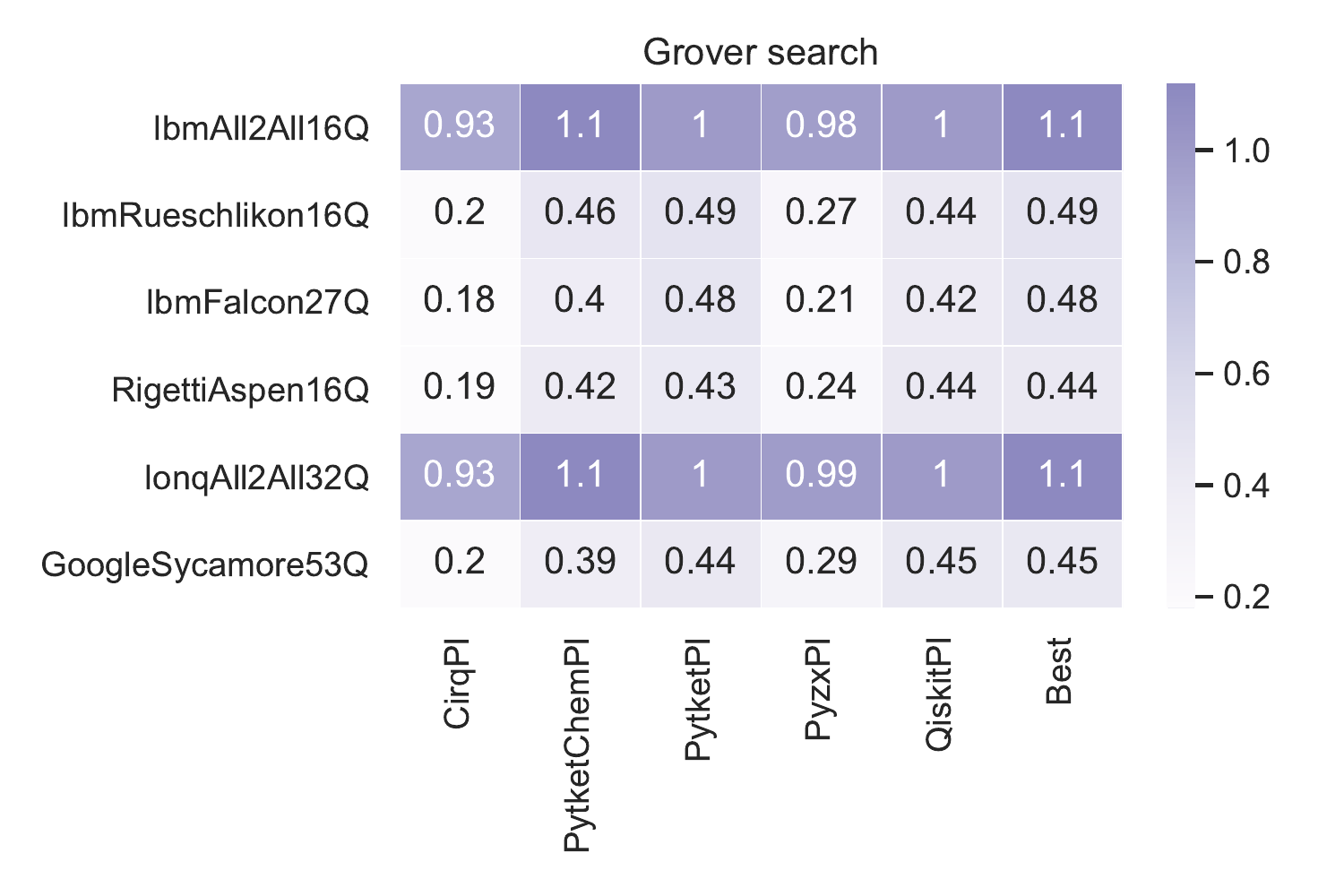}}
        \caption{Cross-hardware comparison of quantum compilation pipelines: two-qubit gates count compression ratio $CF(2Q)$ (higher is better). Final circuits are rebased to hardware native gate sets. The last column (``Best'') shows the maximum compression factor $CF(2Q)$ across all pipelines compiled for a particular hardware architecture.}
        \label{fig:hardware_summary}
    \end{figure}

    In Fig. \ref{fig:hardware_summary} we present two-qubit gate count compression ratios $CF(gc, {2Q})$ for each class of quantum algorithms and each hardware architecture. Overall, circuits compiled for Mock IBM All2All and IonQ architectures show the best compression metrics according to $CF(gc, 2Q)$ and circuit cost function.
    
    Naively one would expect similar results for IBM All2All and IonQ fully connected architectures, while the only difference between two cases is the hardware-native gate set. From Fig. \ref{fig:hardware_summary} we see that the  best values of two-qubit compression factors $CF(gc, 2Q)$ and $CF(depth, 2Q)$ remain the same in both cases, while the single-qubit metrics are significantly worse for IonQ device. The reason is that the compilation pipelines listed in Table \ref{table:pipelines} are not tailored for circuit optimization directly in the gate set of IonQ devices $[XX, \, R_z,\, R_x]$. Instead, they perform circuit optimizations in an intermediate gate set, the output circuit is typically produced in $[CX,\, U_3]$ or $[CZ,\, U_3]$ gate set which is then converted to the IonQ native gate set at the last \textit{Rebase} stage. Gate set rebase is performed by using following identities $CX_{ij}=R_x\left(-\frac{\pi}{2}\right)_i R_z\left(-\frac{\pi}{2}\right)_i R_x\left(-\frac{\pi}{2}\right)_i XX_{ij}\left(-\frac{\pi}{2}\right) R_z\left(\frac{\pi}{2}\right)_i R_x\left(-\frac{\pi}{2}\right)_i R_x\left(-\frac{\pi}{2}\right)_j$, $CZ_{ij}=H_j CX_{ij} H_j$, which inevitably increase single-qubit gate count and hence increases the overall circuit cost function. This problem can be circumvented by performing circuit optimization directly in the IonQ gate set or by performing post-optimizations in the rebased circuit. 
    
    Similarly to IonQ device, when performing compilation for Rigetti Aspen architecture we encounter large depth and total gate count overhead due to contribution of single-qubit gates. Due to the specific  gate set of Rigetti Aspen device that allows only $R_z$ continuous rotations and discrete rotations $R_x(n \pi/2)$, each $U_3$ gate requires 5 hardware-native single-qubit gates to be implemented: $U_3 \sim R_z(\lambda) R_x\left(\frac{\pi}{2}\right) R_z(\theta) R_x\left(-\frac{\pi}{2}\right) R_z(\phi)$. The gate set constraint results in a large  overhead of single-qubit gate operations for Rigetti Aspen device.
    
    Overall,  \textit{QiskitPl}, \textit{PytketPl} and \textit{PytketChemPl} compilation pipelines demonstrated comparable performance, while  \textit{CirqPl} was falling  behind competitors. Interestingly, we found that for UCCSD circuits \textit{PyZXPl} outperformed dedicated \textit{PytketChemPl} pipeline on Mock IBM All2All and IonQ architectures.
    
    When comparing best results with the highest $CF(gc, 2Q)$ among all pipelines, we found  that all architectures with constrained connectivity (excluding Mock IBM All2All and IonQ devices) had similar performance according to $CF(gc, 2Q)$ metric, see Fig. \ref{fig:hardware_summary}. Comparing the maximum values for $CF(gc, 2Q)$ per row in each table (which is denoted as ``Best''), we observe that the deviation of $\max_{pipelines}CF(gc, 2Q)$ across connectivity-constrained architectures is of the order of $10$-$20\%$. We would like to note that the structured circuits used for benchmarking require fewer quantum registers than the number of qubits in the hardware architectures, see Tables  \ref{table:cost_func_params} and \ref{tab:quantum_algos}. Therefore, target circuits are mapped only to a small fraction of available qubits in the case of Google Sycamore and IBM Falcon devices. This explains why in our tests the difference in the two-qubit gate count overhead (for the best performing pipelines) across connectivity-constrained architectures is not significant. 
    
    In this context we would like to point out a recent paper on experimental cross-platform benchmarking of different physical quantum computing devices~\cite{zhu2021cross}. With the development of more advanced cross-hardware benchmarking strategies, they could become an integral part of compiler benchmarks facilitating a more comprehensive cross-testing of various software-hardware co-design architectures. 
    
    \section{Ranking of circuit optimization subroutines}\label{sec:ranking}
    
    In this Section, we compare performance of circuit compression subroutines listed in Table \ref{tab:subroutines_full_list}. We show that the compression power of individual subroutines strongly depends on the class of quantum algorithms considered. However, within the given quantum algorithm class their relative performance remains stable with only small variations between circuit instances.
    
\begin{figure}[H]
        \centering
        \includegraphics[scale=0.7]{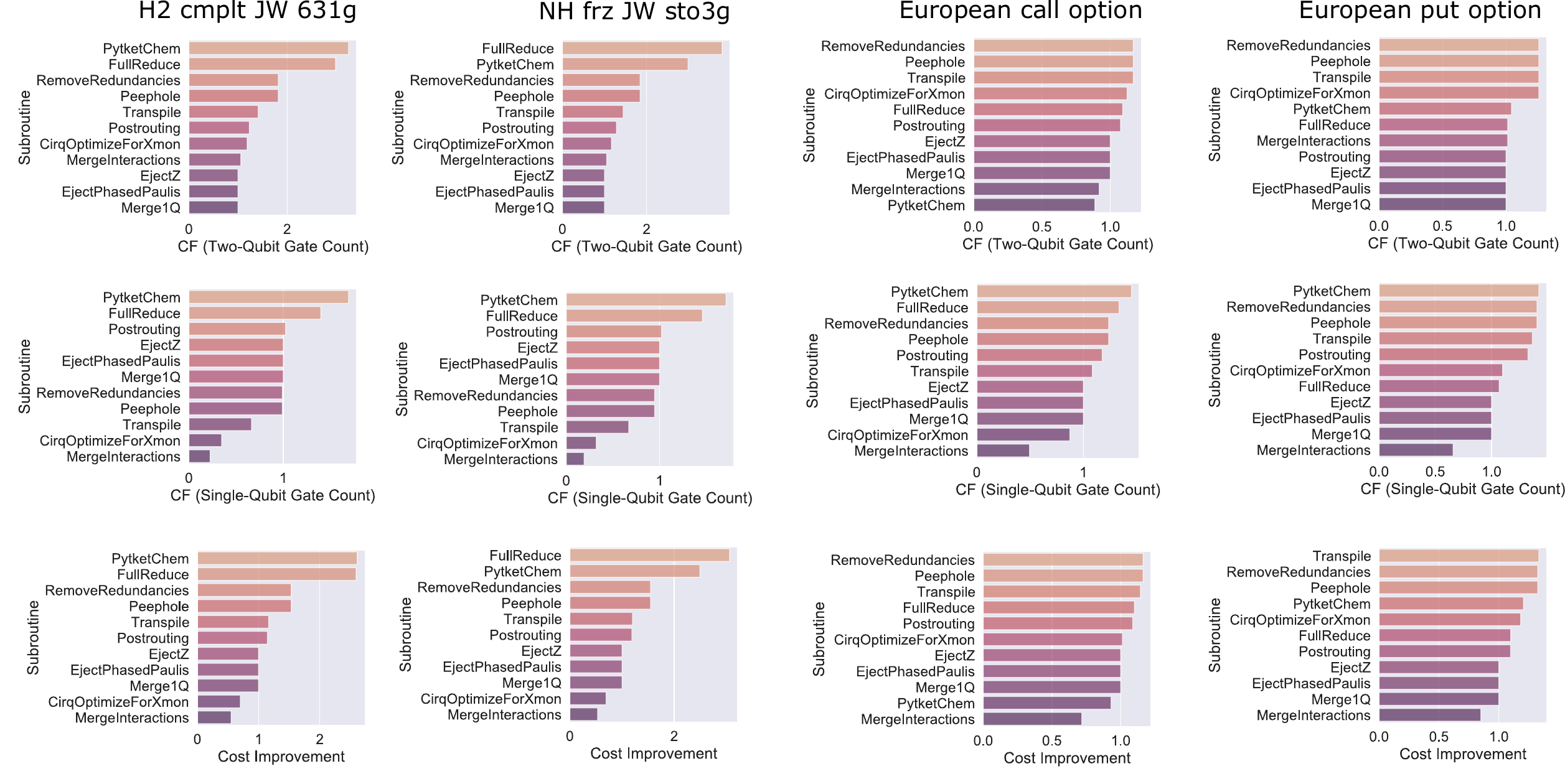}
        \caption{Ranking of compression strategies on architecture with all-to-all connectivity for two classes of target circuits (four circuit instances): the first two columns correspond to UCCSD quantum chemistry circuits: H$_2$ ($8$~qubits) and NH molecules ($10$ qubits), the last two columns correspond to the option pricing quantum algorithm (European call and put option, $11$ qubits). $CF$ stands for compression factor, the cost improvement metric is defined as a ratio of initial to final circuit costs, Eq. \ref{eq:cost_improvement}. }
        \label{fig:ranking}
    \end{figure}

    In Fig. \ref{fig:ranking} we present performance metrics for individual subroutines sorted from the best to the worst performing subroutine. We show results for two classes of quantum circuits: quantum chemistry (two UCCSD circuit instances for H$_2$ and NH molecules, respectively) and quantum algorithms for option pricing (two circuits corresponding to European call and put options). 
    
    For quantum chemistry (UCCSD) circuits, we observe that \textit{PytketChem} and \textit{FullReduce} show the best performance according to $CF(1Q)$ and $CF(2Q)$ compression ratios as the overall circuit cost, which is in agreement with the  results of the Section \ref{sec:structured_circ_compression}. The Cirq's subroutines \textit{MegreInteractions} and \textit{OptimizeForXmon} are the worst performing passes for UCCSD circuits according to the circuit cost metric, even though these subroutines slightly reduced $CX$ count, they resulted in a large $gc(1Q)$ overhead, which pushed the overall circuit cost down.
     
    In the case of option pricing, circuits shown in the third and fourth columns in Fig. \ref{fig:ranking} we found that \textit{RemoveRedundacies}, \textit{Peephole} and \textit{Transpile} are the top-3 performing strategies. \textit{MergeInteractions} increased the output single-qubit gate count, that decreased the overall circuit cost for both call and put option circuit instances. Despite the fact that input circuits within each class have different size (gate counts and number of qubits), the relative ranking of compression subroutines remained roughly the same between the two circuit instances within each target class.
    
\begin{figure}[H]
        \centering
        \includegraphics[scale=0.72]{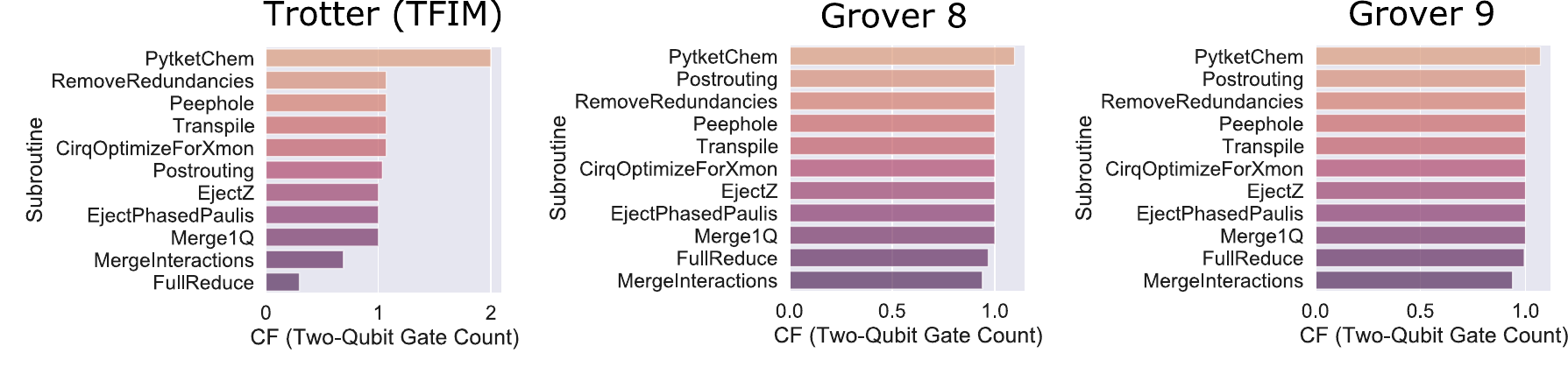}
        \caption{Ranking of performance for individual subroutines according to $CX$ count compression factor (higher is better): Trotter decomposition circuit for 1D transverse field Ising chain model (TFIM) with nearest-neighbour interactions, and Grover search algorithm circuits. The metrics in each plot correspond to a single circuit instance selected from the specified quantum algorithm class. Benchmarking experiment is performed for all-to-all qubit connectivity. }
        \label{fig:ranking_trotter_grover}
    \end{figure}

    In Fig. \ref{fig:ranking_trotter_grover}, we show a similar comparison for Trotter decomposition and Grover search circuits. For the case of Trotter decomposition circuits, subroutines \textit{PytketChem} and \textit{FullReduce} demonstrate the best and the worst two-qubit gates count compression performance, which also matches conclusions of in accordance to Section \ref{sec:structured_circ_compression}. In the case of Grover search circuits only, \textit{PytketChem} subroutine was able to achieve two-qubit gate count compression.
    
    Interestingly, \textit{Merge1Q} subroutine was inactive in all instances (both Fig. \ref{fig:ranking} and \ref{fig:ranking_trotter_grover}), meaning that there were no adjacent single-qubit gates to be merged. This demonstrates that the preliminary rebase to $[CX, R_z, R_x]$ gate set during circuit preprocessing step was efficient.
    
    The relative ranking of subroutines strongly depends on circuit class, by changing circuit class the best performing subroutine can become the worst performing.

    \section{Composite Pipeline}\label{sec:composite}
    
    Quantum circuit optimization algorithms typically consist of a set of elementary subroutines, where each subroutine performs a very specific circuit rewriting task. The exact sequential order of subroutines execution is often chosen heuristically, see e.g. Ref. \cite{nam2018automated}. Hence, it is natural to assume that by rearranging the order of subroutines within a given optimization pipeline one can improve the overall circuit compression performance for the user-specified cost function. Construction of such optimization tailored optimization pipelines for a particular target circuit class requires deep knowledge of strengths/weaknesses of each subroutine and their interaction when executing sequentially, as well as domain knowledge about properties and common patterns in the quantum circuits for a particular class of quantum algorithms.
    
    In this section, we show that by combining compilation subroutines from different modules in a specific way it is possible to achieve better circuit optimization performance, even compared to expert-designed pipelines. We found that the order of compilation subroutines and the choice of the subset of relevant subroutines are dependent on the class of the target quantum circuit. We perform a brute force search of all possible combinations of a selected set of $S$ compression subroutines with a fixed maximum number of stages in the pipeline and utilizing each optimization subroutine only once. 
    
    \begin{figure}[H]
        \centering
        \includegraphics[scale=0.3]{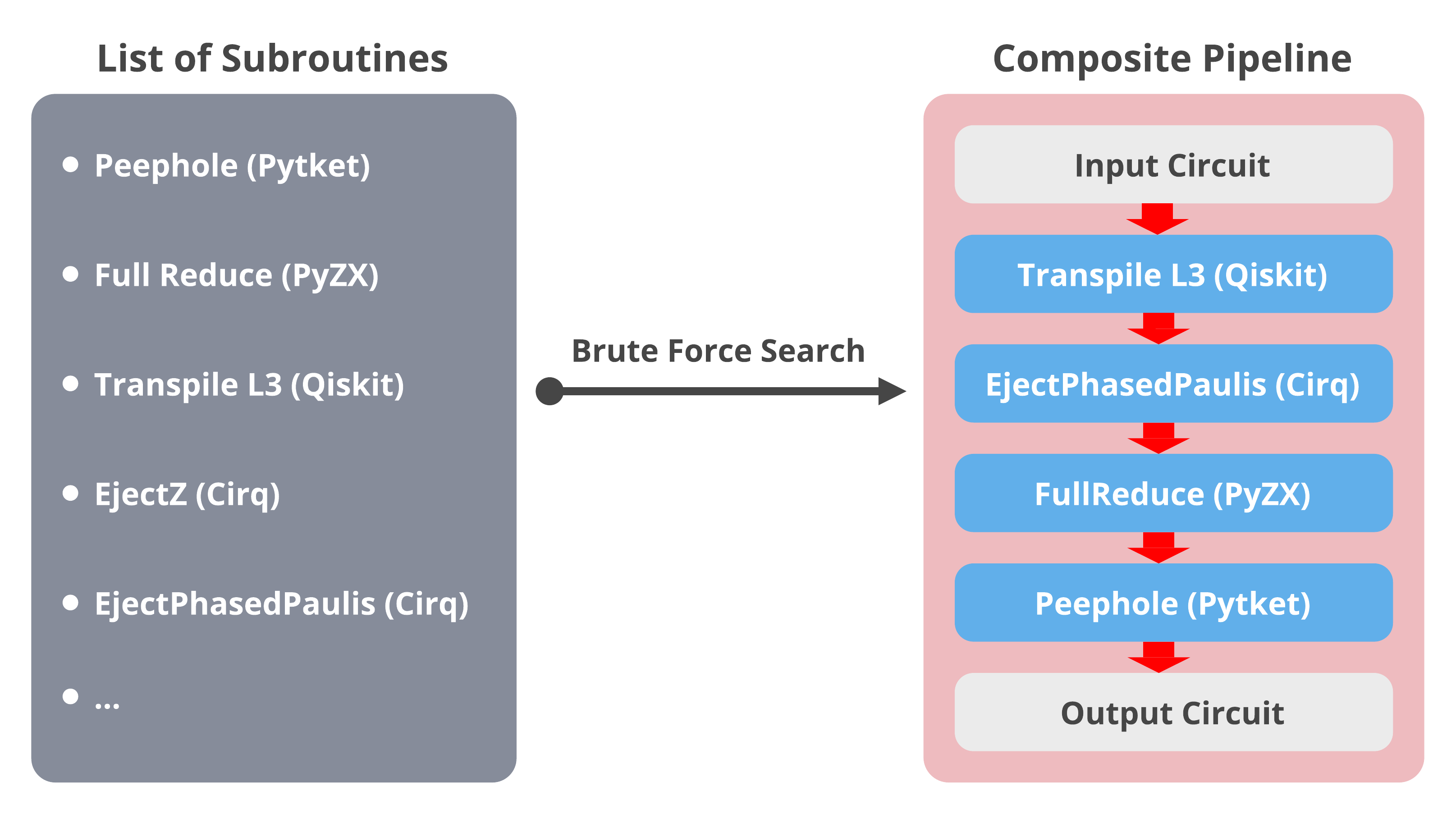}
        \caption{Example of a composite pipeline obtained by brute force search of the best sequence of optimization subroutines.}
        \label{fig:composite_pipeline}
    \end{figure}
    
    We set the maximum exploration depth equal to $MaxD=S$ which corresponds to the maximum number of stages in the pipeline. The total size of the combinatorial search space is $|SearchSpace|=\sum_{n=1}^{MaxD} C^n_{S} n!$, that grows exponentially with the exploration depth $MaxD$. We used the following five circuit compression subroutines for the brute force search of a composite circuit compression pipeline: \textit{EjectPhasedPaulis (Cirq)}, \textit{EjectZ (Cirq)}, \textit{FullReduce (PyZX)}, \textit{Peephole (Pytket)}, \textit{Transpile L3 (Qiskit)}. Due to the exponential scaling of the brute force search complexity, we have to limit ourselves to a small number of trial subroutines. In the case of $S=5$ individual optimization subroutines, the size of the search space is equal to $|SearchSpace|=325$. For simplicity, we assume all-to-all connectivity of qubits, and hence we do not need to consider mapping/routing subroutines. Since some subroutines, in this list have specific requirements for the input circuit gate set (e.g. \textit{FullReduce (PyZX)} does not support $U_3$ gates in the input circuit) we perform gate set rebase to $[CX, R_z, R_x]$ prior to each optimization pass. 
    
    It is worth noting that the chosen subroutines leverage different circuit optimization strategies such as ejection of single-qubit gates towards the end of the circuit, ZX-calculus-based optimization, peephole optimization based on KAK decomposition, utilization of gate cancellation rules. By combining these strategies in a single composite pipeline, it is possible exploit different patterns in the input circuit to achieve a better circuit compression performance.
    
    Table \ref{tab:composite_pip_separate} shows the $CX$ count in the input circuit and in the output circuit after performing circuit optimization with each of the  subroutines from the list. Even though some subroutines such as \textit{EjectPhaseedPaulis} and \textit{EjectZ} do not change the output gate count $gc(CX)$, these subroutines modify the structure of the circuit which can affect the compression performance when executed in conjunction with other subroutines.

    In table \ref{tab:composite_pip_best}, we present the best pipelines we found via brute force search. The pipeline search was performed independently for each circuit. If compared with the data shown in table \ref{tab:composite_pip_separate}, we see that the resulting output circuits have a lower $CX$ gate count compared to the best performing subroutine as well as the dedicated expert-designed pipeline \textit{PytketChemPl} (for UCCSD circuits). Interestingly, we found \textit{EjectZ/EjectPhasedPaulis} subroutines work well in the combination with \textit{FullReduce}, so that \textit{EjectZ/EjectPhasedPaulis}+\textit{FullReduce} can result to an additional $CX$ reduction, when compared to executing \textit{FullReduce} along.
    
    \begin{table}[H]
        \caption{$CX$ count in the optimized circuit after execution of individual subroutines. Highlighted entries show the best (lowest) $gc(CX)$ for each of the four input circuits. We assume all-to-all qubit connectivity, final circuits are rebased to $[CX, U_3]$ gate set.}
        \centering
        \resizebox{\textwidth}{!}{\begin{tabular}{|l|c|c|c|c|}
            \hline
            \multicolumn{1}{|c|}{\textbf{Stage}} & \textbf{H2cmpltJW631g (8Q)} & \textbf{NHfrzJWsto3g (10Q)} & \textbf{Call Option (11Q)} & \textbf{Put Option (11Q)} \\
            \hline
            \textbf{Input circuit} & 768 & 3896 & 229 & 73\\
            \hline
            Peephole (Pytket) & 422 & 2108 & \textbf{196} & 58 \\
            \hline
            EjectPhasedPaulis (Cirq) & 768 & 3896 & 229 & 73 \\
            \hline
            EjectZ (Cirq) & 768 & 3896 & 229 & 73\\
            \hline
            FullReduce (PyZX) & 242 & \textbf{1051} & 215 & 76 \\
            \hline
            Transpile (Qiskit) & 544 & 2698 & \textbf{196} & \textbf{58} \\
            \hline
            ChemPl (Pytket)  & \textbf{236} & 1305 & 258 & 70\\
            \hline
        \end{tabular}}
        \label{tab:composite_pip_separate}
    \end{table}

    \begin{table}[H]
        \caption{The best composite pipelines found via brute force search by combining optimization subroutines from Table \ref{tab:composite_pip_separate} with the maximum exploration depth (maximum number of allowed subroutines) $MaxD=5$.}
        \centering
        \begin{tabular}{|l|c|c|c|c|}
            \hline
            \multicolumn{1}{|c|}{\textbf{Metrics}} & \textbf{H2cmpltJW631g (8Q)} & \textbf{NHfrzJWsto3g (10Q)} & \textbf{Call Option (11Q)} & \textbf{Put Option (11Q)}\\
            \hline
            \makecell{\textbf{The best pipeline} \\ Stage order $\downarrow$} 
            & \makecell{Transpile (Qiskit) \\ FullReduce (PyZX) \\ Peephole (Pytket)} & \makecell{Transpile (Qiskit) \\ FullReduce (PyZX) \\ Peephole (Pytket)} & \makecell{Transpile (Qiskit) \\ EjectPhasedPaulis (Cirq) \\ FullReduce (PyZX) \\ Peephole (Pytket)} & \makecell{EjectPhasedPaulis (Cirq) \\ FullReduce (PyZX) \\ Peephole (Pytket) \\ Transpile (Qiskit) } \\
            \hline
            \textbf{Final $CX$ count} & \textbf{206} & \textbf{862} & \textbf{147} & \textbf{54}\\
            \hline
        \end{tabular}
        \label{tab:composite_pip_best}
    \end{table}

    \section{Conclusions and outlook}\label{sec:conclusions}

    In this paper, we presented an open-source software platform Arline Benchmarks for automated benchmarking of various subroutines for quantum circuit optimization. We developed benchmarking methodologies and performed a systematic analysis of circuit optimization as well as qubit routing/mapping subroutines of compilation frameworks. In order to provide a fair comparison between compilers, the input circuits were translated to a fixed gate set and output circuits were translated to a hardware native gate set. Besides  standard metrics used for benchmarking of quantum compilers, we considered an aggregate metric characterizing circuit quality in the NISQ regime -- circuit cost function. The phenomenological circuit cost function is evaluated based on the gate content of the circuit and quantum hardware parameters, such as fidelities of single-qubit and two-qubit gates. Extension of compiler benchmarking functionality for the fault-tolerant regime could be added in future releases of Arline Benchmarks.

    Using Arline Benchmarks platform, we performed a comprehensive analysis
    of quantum compilers performance on random and structured circuits  corresponding to quantum algorithms.
    We compared performance of qubit routing subroutines on two types of qubit connectivities: random k-regular graphs and popular hardware architectures of real devices. 
    By performing benchmarking of circuit optimization pipelines for structured target circuits we showed that performance of subroutines tailored for a specific circuit class/quantum algorithm (e.g. UCCSD circuits) are sensitive to specific gate patterns in the input circuit. Introducing randomization of gate placement in these circuits significantly impacts compression rates. 

    
    
    By leveraging the cross-compiler functionality of Arline Benchmarks, we explored the idea of a composite circuit optimization pipeline. We showed that by stacking subroutines from different vendors in a specific target-dependent sequence it is possible to achieve higher circuit compression rates. We demonstrated that the sequential order of execution of subroutines in the pipeline is crucial, and by performing a brute force search within a restricted space of candidate pipelines, we were able to find improved pipelines even compared to expert-designed solutions.

    In future releases of Arline Benchmarks we are planning to extend the list of supported frameworks, by integrating with Quilc \cite{quilc2020}, ProjectQ \cite{steiger2018projectq}, Staq \cite{amy2020staq} and VOQC \cite{hietala2021verified}. In addition, we plan to expand the database of target circuits/quantum algorithms, set of metrics such a $T$ gate count, quantum volume for compressed circuits,  as well as include other common test types such as unitary synthesis and quantum state preparation.

    \bibliography{papers}

\end{document}